\documentclass[aps,prd,showpacs,floatfix,nofootinbib,amssymb,flushbottom,preprint]{revtex4-1}
\usepackage{graphicx}
\usepackage{slashed}
\usepackage{booktabs}
\usepackage{latexsym}
\usepackage{amsmath}
\usepackage{tabularx}
\usepackage{dsfont}
\usepackage[utf8]{inputenc}
\usepackage[a4paper]{geometry}
\usepackage[T1]{fontenc}
\usepackage{textcomp}
\usepackage{latexsym}
\newcommand{\nn}{\nonumber}
\newcommand{\be}{\begin{equation}}
\newcommand{\ee}{\end{equation}}
\newcommand{\bea}{\begin{eqnarray}}
\newcommand{\eea}{\end{eqnarray}}
\newcommand{\al}{\alpha}
\newcommand{\als}{\alpha_{\mathrm{s}}}

\newcommand{\MS}{\overline{\mathrm{MS}}}
\newcommand{\lQ}{\Lambda_{\mathrm{QCD}}}
\newcommand{\Eqre}{Eq.~\protect\eqref}

\newcommand{\nmax}{{n_\mathrm{max}}}
\newcommand{\NSPT}{\mathrm{NSPT}}
\newcommand{\MC}{\mathrm{MC}}
\newcommand{\NP}{\mathrm{NP}}
\newcommand{\G}{\mathrm{G}}
\newcommand{\latt}{\mathrm{latt}}
\newcommand{\pert}{\mathrm{pert}}
\newcommand{\soft}{\mathrm{soft}}

\newcommand{\order}{\mathcal{O}}
\newcommand{\ra}[1]{\renewcommand{\arraystretch}{#1}}
\newcommand{\chired}{\chi^2_{\mathrm{red}}}
\newcommand{\PBC}{\mathrm{PBC}}
\newcommand{\TBC}{\mathrm{TBC}}

\begin{document}
\title{\boldmath
Perturbative expansion of the plaquette to $\order(\al^{35})$  
in four-dimensional SU(3) gauge theory 
\unboldmath}
\author{Gunnar S.\ Bali}
\email{gunnar.bali@ur.de}
\affiliation{Institut f\"ur Theoretische Physik, Universit\"at
Regensburg, D-93040 Regensburg, Germany}
\affiliation{Tata Institute of Fundamental Research, Homi Bhabha Road, Mumbai 400005, India}
\author{Clemens Bauer}
\affiliation{Institut f\"ur Theoretische Physik, Universit\"at
Regensburg, D-93040 Regensburg, Germany}
\author{Antonio Pineda}
\email{AntonioMiguel.Pineda@uab.es}
\affiliation{Grup de F\'{\i}sica Te\`orica, Universitat
Aut\`onoma de Barcelona, E-08193 Bellaterra, Barcelona, Spain}

\date{\today}

\begin{abstract}
\noindent
Using numerical stochastic perturbation theory, we determine the first 35 infinite volume coefficients of the perturbative
expansion  in powers
of the strong coupling constant $\al$ of the plaquette  in $\mathrm{SU}(3)$ gluodynamics.
These coefficients are obtained in lattice regularization with the standard
Wilson gauge action. The on-set of the dominance of the
dimension four renormalon associated
to the gluon condensate is clearly observed. We determine the
normalization of the corresponding singularity in the Borel plane and
convert this into the $\MS$ scheme. We also comment on
the impact of the renormalon on non-perturbative determinations of 
the gluon condensate.
\end{abstract}
\maketitle

\section{Introduction}
Perturbative expansions,
$\sum_{n=0}^{\nmax}a_n\al^{n+1}$, in powers of the coupling parameter
$\al$ of four dimensional non-Abelian
gauge theories are expected to be divergent as $\nmax\rightarrow\infty$.
The structure of the operator product expansion (OPE)
determines particular patterns of
asymptotic divergence that are usually
named renormalons~\cite{Hooft}. 

In three recent articles~\cite{Bauer:2011ws,Bali:2013pla,Bali:2013qla},
we presented compelling evidence for the existence
of the leading renormalon associated to the (dimension one) pole mass
of heavy quark effective theory (or potential non-relativistic QCD),
as expected from the standard OPE \cite{Bigi:1994em,Beneke:1994sw}. This was achieved by
expanding the energy of a static source in a lattice scheme
to $\order(\al^{20})$ using numerical stochastic perturbation
theory (NSPT)~\cite{DRMMOLatt94,DRMMO94}. For a review of NSPT,
see Ref.~\cite{DR0}.
As a by-product, the normalization of this singularity in the Borel plane
was obtained and converted into the modified minimal subtraction ($\MS$) scheme.

The situation regarding the renormalon associated with the
(dimension four) gluon condensate~\cite{Vainshtein:1978wd} is less
well settled. This condensate determines the leading non-perturbative correction, e.g., to the QCD Adler function, or, in lattice regularization, to the 
plaquette.
Previously, diagrammatic~\cite{DiGiacomo:1981wt,Alles:1993dn}
and several
high-order NSPT computations~\cite{DiRenzo:1995qc,Burgio:1997hc,DiRenzo:2000ua,Rakow:2005yn,Horsley:2012ra}
of the plaquette 
have been carried out in lattice regularization,
with conflicting conclusions
regarding the convergence properties and the position
of the leading singularity in the Borel plane.

The position and normalization of this singularity and
the value of the gluon condensate are not only topics of
theoretical debate but also
impact on important questions of
particle physics phenomenology. For instance, precision determinations 
of the strong coupling constant $\als$ from $\tau$-meson decays 
rely on perturbative series that are also sensitive to the 
gluon condensate renormalon~\cite{Beneke:2008ad,Pich:2013sqa}.
The same applies to computations of partial decay rates of
a Higgs particle into heavy quark-antiquark pairs, see e.g.~Ref.~\cite{Broadhurst:2000yc}.
From the theoretical side, high-order perturbative
series in quantum mechanical systems \cite{ZinnJustin:2004ib,Basar:2013eka} and quantum field theories~\cite{Dunne:2012ae,Aniceto:2013fka,Cherman:2013yfa}
have recently been studied in the framework of
resurgent trans-series.
The relevance of this promising work to renormalons in QCD
has yet to be elucidated.

The order in $\al$ at which the renormalon dominates the 
asymptotic behaviour of the perturbative series is proportional
to the dimension of the associated operator.
In our recent investigation of an infrared renormalon associated to
a dimension one operator~\cite{Bauer:2011ws,Bali:2013pla},
the on-set of the asymptotic
behaviour in the (Wilson) lattice scheme was observed at
orders $\sim$ 7~--~9 in $\al$. 
Hence, in the case of the dimension four gluon condensate, 
the order of the expansion
necessary to enable detection of the corresponding renormalon
needs to be multiplied by a factor of approximately four.
Previous computations of the plaquette
in the Wilson lattice scheme, however, have only been carried out
up to $\order(\al^{20})$
in the strong coupling constant~\cite{Horsley:2012ra}.
In this case no
volume was larger than $12^4$. For
volumes of $24^4$ points previous results only exist
up to $\order(\al^{10})$~\cite{DiRenzo:2000ua},
and for $32^4$  up to $\order(\al^{3})$~\cite{DiRenzo:2004ge}.

A controlled study of the asymptotic behaviour of the series
and of the normalization of the renormalon
is required to determine the gluon condensate and
its intrinsic ambiguity. This application and its phenomenological
impact will be discussed
in a forthcoming paper. Here we
concentrate on the technical details of our simulations and, in particular,
on the determination of the infinite volume coefficients to $\order(\al^{35})$
from NSPT simulations of finite volumes of up to $40^4$ sites. In spite of
several optimizations, the computer time and memory requirements were
considerable. For instance, the storage of two copies of a $40^4$ lattice to
order $\al^{30}$ alone requires about 170 GBytes of main memory,
clearly necessitating
the use of parallel systems.

This article is organized as follows. In Sec.~\ref{sec:simul}
we introduce our notation, the action, the lattice volumes
and the simulation methods used.
In Sec.~\ref{sec:fse} we discuss the dependence of the coefficients of the perturbative series of the plaquette
on the volume and boundary conditions.
In Sec.~\ref{sec:fits} we extrapolate these coefficients to infinite volume. 
Finally, in Sec.~\ref{sec:Renormalons} we compare these infinite
volume results
against renormalon-based expectations for their high-order behaviour,
determine the normalization of the gluon condensate renormalon and 
discuss the impact of its value on non-perturbative determinations of 
the gluon condensate itself, before we
conclude.

\section{Simulation details}
\label{sec:simul}
We introduce some of our notations and list the simulated lattice volumes.
We also explain how we account for errors associated to
finite Langevin time steps and qualitatively survey the volume dependence
of our results.
We refer to Ref.~\cite{Bali:2013pla} for a more detailed account of
the theoretical and numerical methods used, their implementation
and tests.
\subsection{Notation and simulated volumes}
We study hypercubic Euclidean spacetime
lattices $\Lambda_E$ with a lattice spacing $a$
and $N^4$ sites, labelled
by $x=am\in\Lambda_E$, $m=(m_{\mu})=(m_1,m_2,m_3,m_4)$, $m_{\mu}=0,\ldots,N-1$.
We realize linear dimensions $N\leq 40$, twisted boundary
conditions (TBC)~\cite{'tHooft:1979uj} in
all three spatial directions
$\mu=1,2,3$, and
periodic boundary conditions in time $\mu=4$
as, e.g., detailed in Ref.~\cite{Bali:2013pla}.

We employ the standard Wilson gauge action
\begin{equation}
\label{eq:wilson}
S = \beta\sum_{\substack{x\in\Lambda_E\\ \mu>\nu}}
P_{x,\mu\nu}=\int\!d^4x\,\sum_{\mu,\nu,c}\frac{1}{4}G_{\mu\nu}^c(x)G_{\mu\nu}^c(x)\times
\left[
1+\order(a^2)\right]\,,
\end{equation}
where
$\beta=6/g^2=3/(2\pi\al)$ and $\al=g^2/(4\pi) \equiv \al(a^{-1})$ is the bare lattice coupling. $c=1,\ldots,8$ is the adjoint colour index and
\begin{equation}
P_{x,\mu\nu}=1-\frac{1}{6}\mathrm{Tr}\left(U_{x,\mu\nu}+U_{x,\mu\nu}^{\dagger}\right)\,.
\end{equation}
$U_{x,\mu\nu}$ denotes the oriented product of four link
variables
\be
\label{eq:linkvar}
U_{x,\mu}=
\mathcal{P}\exp\left[ig\int_x^{x+a\hat{\mu}}\!dx'_{\mu}\,A_{\mu}(x')\right]
\approx e^{igaA_{\mu}[x+(a/2)\hat{\mu}]}\in\mathrm{SU}(3)\,,
\ee
enclosing the elementary square (plaquette) with corner positions
$x$, $x+a\hat{\mu}$, $x+a(\hat{\mu}+\hat{\nu})$ and $x+a\hat{\nu}$.
$\mathcal{P}$ denotes path ordering and $A_{\mu}=A_{\mu}^ct^c$ as usual.
Note that, using the above normalization convention
for the action, the gluonic field strength tensor reads
\begin{equation}
G_{\mu\nu}=-\frac{i}{g}[D_{\mu},D_{\nu}]=\partial_{\mu}A_{\nu}
-\partial_{\nu}A_{\mu}+ig[A_{\mu},A_{\nu}]\,.
\end{equation}

We define the vacuum expectation value of a generic operator $B$ of
engineering dimension zero as
\begin{equation}
\label{eq:ANP}
\langle B\rangle \equiv \langle \Omega |B|\Omega\rangle =\frac{1}{Z}\int\![dU_{x,\mu}]\,e^{-S[U]} B[U]
\end{equation}
with the partition function $Z=\int[dU_{x,\mu}]\,e^{-S[U]}$
and measure $[dU_{x,\mu}]=\prod_{x\in\Lambda_E,\mu}dU_{x,\mu}$.
$|\Omega\rangle$ denotes the vacuum state.
$\langle B\rangle$ will depend on the lattice extent $Na$ and
spacing $a$.
The  coefficients $b_n$ of its perturbative expansion
\begin{equation}
\label{eq:A}
\langle B \rangle_{\pert}(N) \equiv \frac{1}{Z}\left.\int\![dU_{x,\mu}]\,e^{-S[U]} B[U]\right|_{\NSPT}
=\sum_{n\geq 0}b_n(N)\al^{n+1}
\end{equation}
are obtained by Taylor expanding the link variables
$U_{x,\mu}$ of \Eqre{eq:linkvar}
in powers of $g$ before averaging over the gauge configurations
by means of a Langevin simulation with
a time step $\epsilon>0$
(NSPT)~\cite{DRMMOLatt94,DRMMO94,DR0}. 

In \Eqre{eq:A} we have made explicit that the
coefficients $b_n$ are functions of the linear lattice size
$N$. However, we emphasize that the $b_n(N)$
do not depend on the lattice spacing $a$:
the above integration is
over the dimensionless link variables $U_{x,\mu}$
and $a$ can be absorbed into the definition of
the $A_{\mu}(x)$ fields of \Eqre{eq:linkvar}.

The integration over the gauge variables in \Eqre{eq:A} is finite
for all non-zero modes but divergent for the zero modes (see,
for instance, the discussion in
Ref.~\cite{pbcproblem}). Perturbation theory in lattice
regularization with TBC eliminates zero
modes~\cite{Coste:1985mn,Luscher:1985wf}, 
yielding finite, well-defined results for the coefficients
$b_n$. This is not the case
for periodic boundary conditions (PBC) where
zero modes are usually subtracted ``by hand'' to give finite results.
We will see in Secs.~\ref{sec:PBC} and \ref{sec:fitPBC} that this
causes some problems. 

\begin{table}
\caption{\it 
Maximal order of the plaquette expansion and
respective linear lattice extent $N$.
In total, we have considered 21 different volumes. 
Volumes for which $\epsilon\rightarrow 0$ extrapolations
in the Langevin time step were carried out are labelled by
bracketed bold superscripts that indicate the maximal order to which
$\epsilon=0$ results are available.
For the remaining lattices only a single value
$\epsilon=0.05$ was realized.\\~
}
\label{tab:volumesPlaq}
\begin{ruledtabular}
\begin{tabular}{>{$}c<{$}|>{$}c<{$}}
\mathrm{order} & N\\\hline
\order\left(\alpha^{5}\right)  & 11, 13\\
\order\left(\alpha^{20}\right) & 14\\
\order\left(\alpha^{30}\right) & 12, 40\\
\order\left(\alpha^{35}\right) & 3, 4^{(\mathbf{5})}, 5, 6^{(\mathbf{12})}, 9, 10^{(\mathbf{12})}, 28^{(\mathbf{35})}, 30\\  
\order\left(\alpha^{40}\right) & 7, 8^{(\mathbf{12})}, 16^{(\mathbf{12})}, 18, 20, 22, 24, 32
\end{tabular}
\end{ruledtabular}
\end{table}

We define 
\begin{equation}
\label{eq:plaqnorm}
P_x=\frac{1}{6}\sum_{\mu>\nu}P_{x,\mu\nu}=a^4\frac{\pi\alpha}{9}
\frac{1}{4}G_{\mu\nu}^c(x)G_{\mu\nu}^c(x) + \order(a^6)\,.
\end{equation}
The average plaquette
\begin{equation}
\langle P\rangle=\langle P_0\rangle=\frac{1}{N^4}\sum_{x\in\Lambda_E}\langle P_x\rangle
\end{equation}
does not depend on the spacetime point, due to translational
invariance of expectation values, and hence we drop its position index.
In this article we compute its expansion coefficients $p_n(N)$,
\begin{equation}
\label{eq:expand}
\langle P \rangle_{\pert}(N)=\sum_{n\geq 0}p_n(N)\al^{n+1}
\,,
\end{equation}
for the volumes and up to the
maximal orders
in $\al$ displayed in Table~\ref{tab:volumesPlaq}.
Due to increases of statistical
errors and autocorrelation times at very high orders,
we decided to restrict ourselves to
$\nmax+1\leq 35$ in our final analysis.

\subsection{Simulations and extrapolation to a vanishing Langevin time step}
\label{sec:langevin}
In our simulation the second-order integrator
introduced in Ref.~\cite{Torrero:2008vi}
and detailed in Ref.~\cite{Bali:2013pla} is employed.
We use stochastic gauge fixing to avoid
run-away trajectories, see e.g.~Ref.~\cite{DR0},
and thermalize each order $j-1$, before ``switching
on'' the next order $j\leq n$. After the
thermalization phase,
``measurements'' are taken
and analysed following Ref.~\cite{Wolff:2003sm} for the treatment of
(auto-)correlations.

\begin{figure}
\begin{center}
\includegraphics[width=0.75\textwidth]{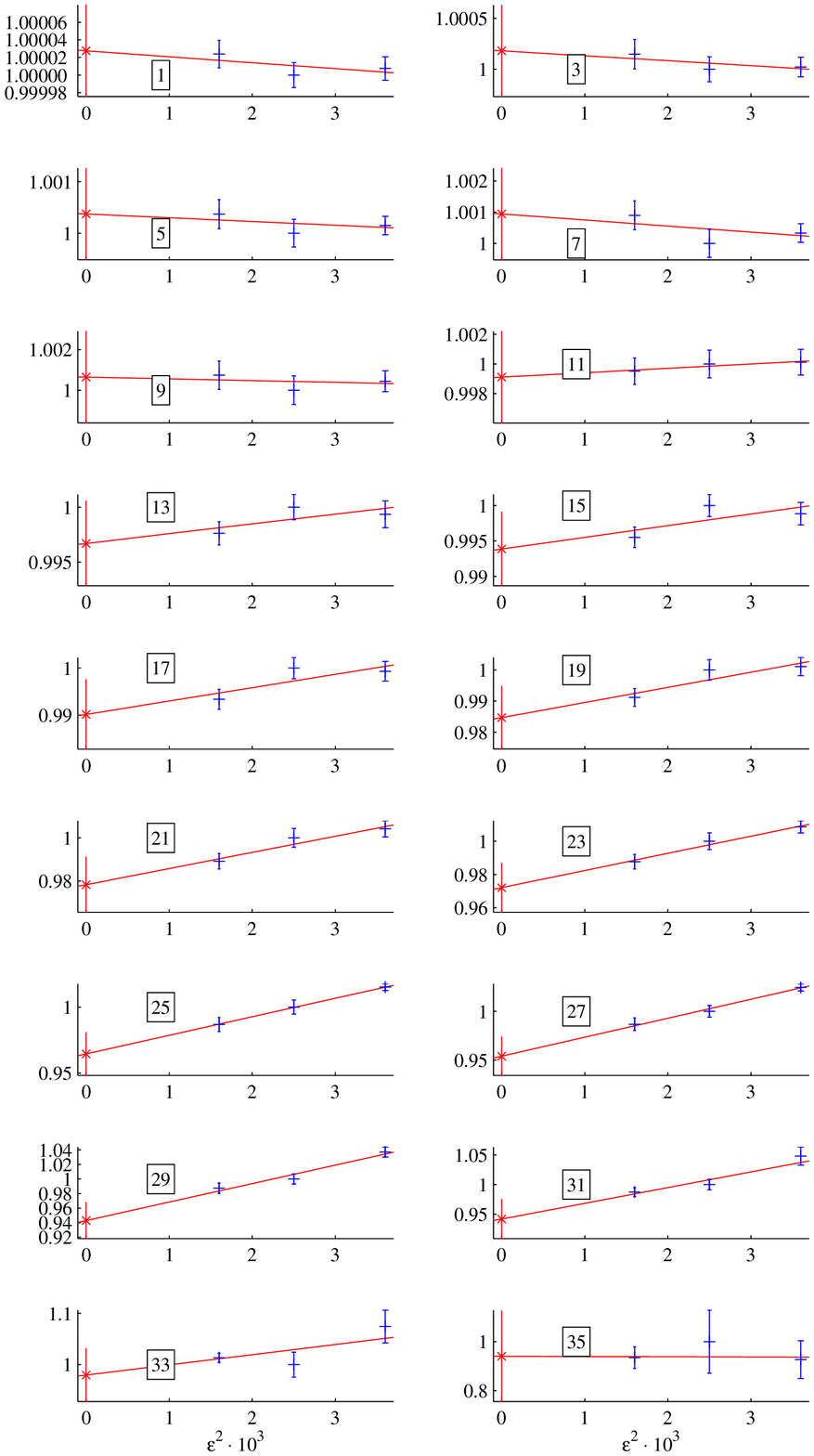}\\[-1.5cm]~
\end{center}
\caption{\it  Time step extrapolations of $p_n(N\!\!=\!\!28;\epsilon)/p_n(N\!\!=\!\!28;0.05)$.
Boxed numbers refer to the order in $\alpha$:
$n+1=1, 3,\ldots, 35$. The left-most
symbols are the extrapolated values.
}
\label{fig:PlaqL28}
\end{figure}

Due to issues of numerical stability and the expense of
generating a sufficiently large number of effectively
statistically independent measurements, the time step
$\epsilon$ cannot be taken arbitrarily small. We carry out most simulations 
at $\epsilon=0.05$. However, we investigate the
$\order(\epsilon^2)$ discretization errors by additionally simulating
$\epsilon=0.04$ and
0.06 on the $N=4,6, 8, 10, 16$ and 28 lattices to the maximal order in $\al$ stated in Table~\ref{tab:volumesPlaq}.

We show the $\epsilon^2\rightarrow 0$ extrapolation of the
$N=28$ data 
in Fig.~\ref{fig:PlaqL28} for the example
of odd orders $n+1$.
For orders $n+1\leq 15$ no statistically significant
slopes  can be detected
and the $\epsilon=0.05$ results are in perfect agreement
within errors
with the $\epsilon\rightarrow 0$  extrapolations. 
(One notable exception is the
$\order(\al^{2})$ data, not depicted here.) 
For higher
orders the non-vanishing size of $\epsilon$ introduces errors,
which we estimate in the following way. From the $N=28$ data we compute
the relative difference between the value of a coefficient
$p_n$ obtained at the finite value $\epsilon=0.05$ and the
extrapolated result:
\begin{equation}
d_n=\left|1-\frac{p_n(\epsilon=0)}{p_n(\epsilon=0.05)}\right|
\,.
\end{equation}
For all the volumes and orders
where no $\epsilon\rightarrow 0$ extrapolation was carried out, 
we use $d_np_n$ as the estimate of the uncertainty due to the
non-zero time step.
We then add $d_np_n$ to the
respective statistical error of $p_n$ obtained at $\epsilon = 0.05$
in quadrature. 
For the coefficients $p_n(N)$ where the $\epsilon$-extrapolation
has been carried out, we use 
the extrapolated value $p_n(N;\epsilon=0)$ and the associated error of the $\epsilon$-extrapolation instead. 

\begin{figure}
\begin{center}
\includegraphics[width=0.9\textwidth]{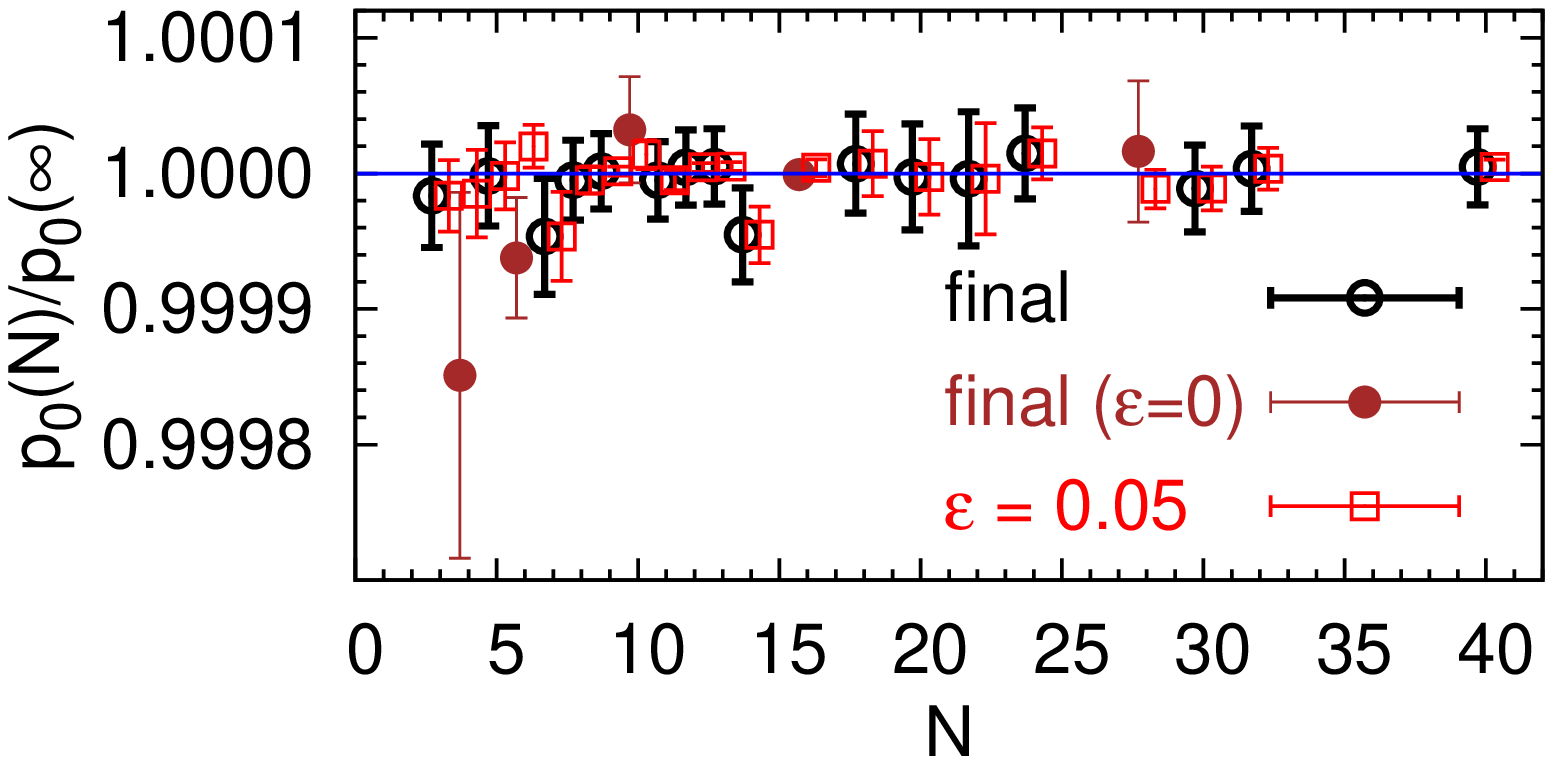}\\
\includegraphics[width=0.9\textwidth]{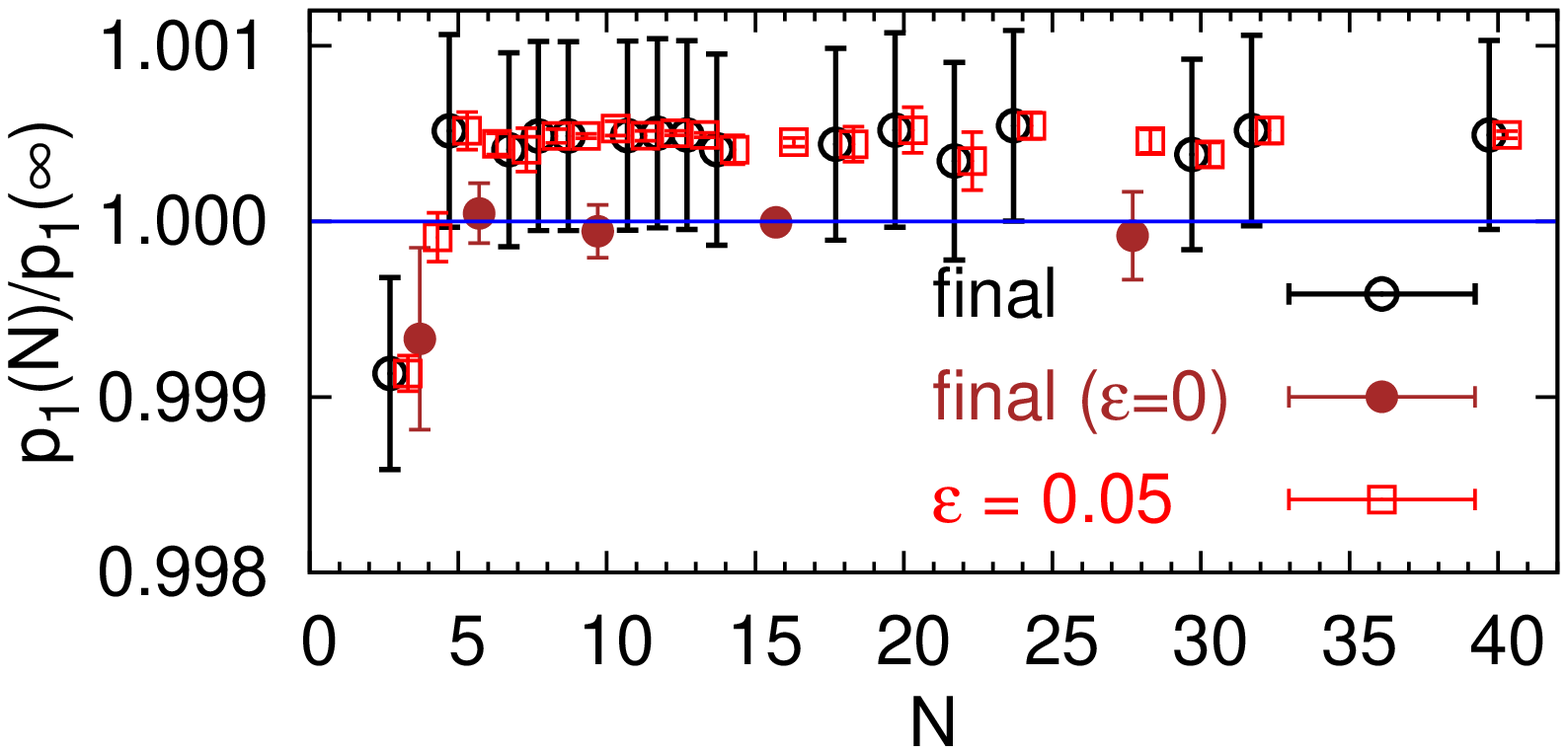}
\end{center}
\caption{\it The coefficients $p_0(N)$
(upper panel)
and $p_1(N)$ (lower panel) for different
linear lattice extents $N$,
normalized with respect to the
infinite volume
expectations from diagrammatic perturbation theory.
Circles denote the final values
obtained either by increasing the
respective errors (empty circles) or by extrapolating
to $\epsilon=0$ (full circles) as detailed in the text. Squares 
correspond to the values obtained at the fixed time step $\epsilon=0.05$.
For clarity the symbols have been
shifted horizontally by different off-sets.}
\label{fig:DATA_p0}
\end{figure}

\begin{figure}
\begin{center}
\includegraphics[width=0.9\textwidth]{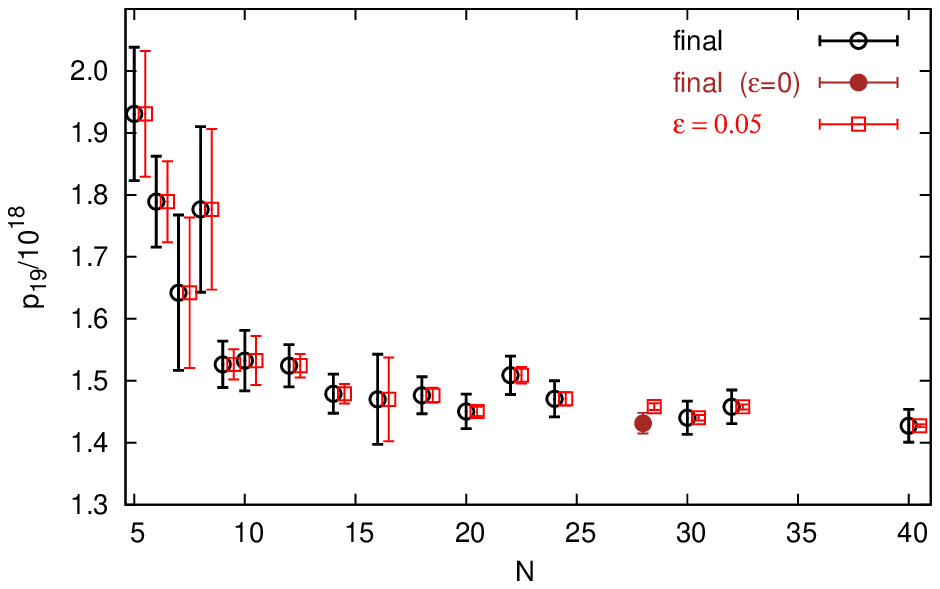}
\end{center}
\caption{\it The coefficient $p_{19}(N)$
versus the linear lattice size $N$. 
The full circle denotes the $\epsilon=0$ extrapolated result,
which at this order is only available for $N=28$.
Open circles are the ``final'' values obtained by 
increasing the errors as detailed in the text, squares
are the results obtained at a
fixed time step $\epsilon=0.05$. The symbols have been
shifted horizontally to enhance readability.}
\label{fig:DATA}
\end{figure}

Figs.~\ref{fig:DATA_p0} and \ref{fig:DATA} show the impact of the
$\epsilon$-extrapolation error on $p_{0,1,19}(N)$. 
In the upper panel of Fig.~\ref{fig:DATA_p0} we normalize the data to
the analytically known value $p_0(\infty)=p_0(N)=4\pi/3$.
We observe perfect agreement with this
expectation. The $\epsilon$-extrapolation errors
dominate for large volumes where the statistical errors
are small.
This is a general tendency for all orders $n$, but more pronounced for
large $n$-values, see Fig.~\ref{fig:DATA}. In the lower panel
of Fig.~\ref{fig:DATA_p0} we normalize the data to the known
value $p_1(\infty)$. This plot further
illustrates the quality of the 
$\epsilon$-extrapolation and that our error estimates
are reasonable. Note that in this case a non-zero slope of the
$\epsilon^2$-extrapolation was detected. For all but one of the volumes
for which the extrapolation in $\epsilon^2$ was performed
($N=6, 8, 10, 16, 28$) we find
perfect agreement within small errors with the infinite
volume result.
Only for $N=4$ are finite size effects significant.
We also see how our procedure to estimate the $\epsilon$-extrapolation
error (based on the deviation at $N=28$) correctly captures the
systematics for all the volumes for which we could not perform
an $\epsilon$-extrapolation. 

Since the gauge action and the algorithm are
local in spacetime and Langevin time one may expect the
$\epsilon^2$-slopes to become independent of $N$ for
sufficiently large lattice extents $N$, with $1/N^2$ corrections
that will depend on the order of the expansion.
Indeed, this expectation seems to be supported by our data, see
Fig.~\ref{fig:DATA_p0}, where the shifts between the $\epsilon=0.05$
and extrapolated data are similar in sign and magnitude for all volumes.  
However, in the present article we try to inject
as little prejudice as possible into the analysis.
Therefore, we follow the more conservative approach outlined above and
abstain from using this information in the $\epsilon$-extrapolation.

\subsection{Qualitative survey of PBC and TBC results}
\label{sec:quali}
In our simulations we realize TBC. Numerically, these
boundary conditions have the advantage of reduced
statistical fluctuations and smaller autocorrelation
times, due to the complete absence of
zero momentum modes. Moreover, at small orders, these
boundary conditions reduce finite size effects, and
--- as we shall see below --- we can theoretically control
TBC volume effects much better than PBC ones.

\begin{figure}
\centerline{\includegraphics[width=0.95\textwidth]{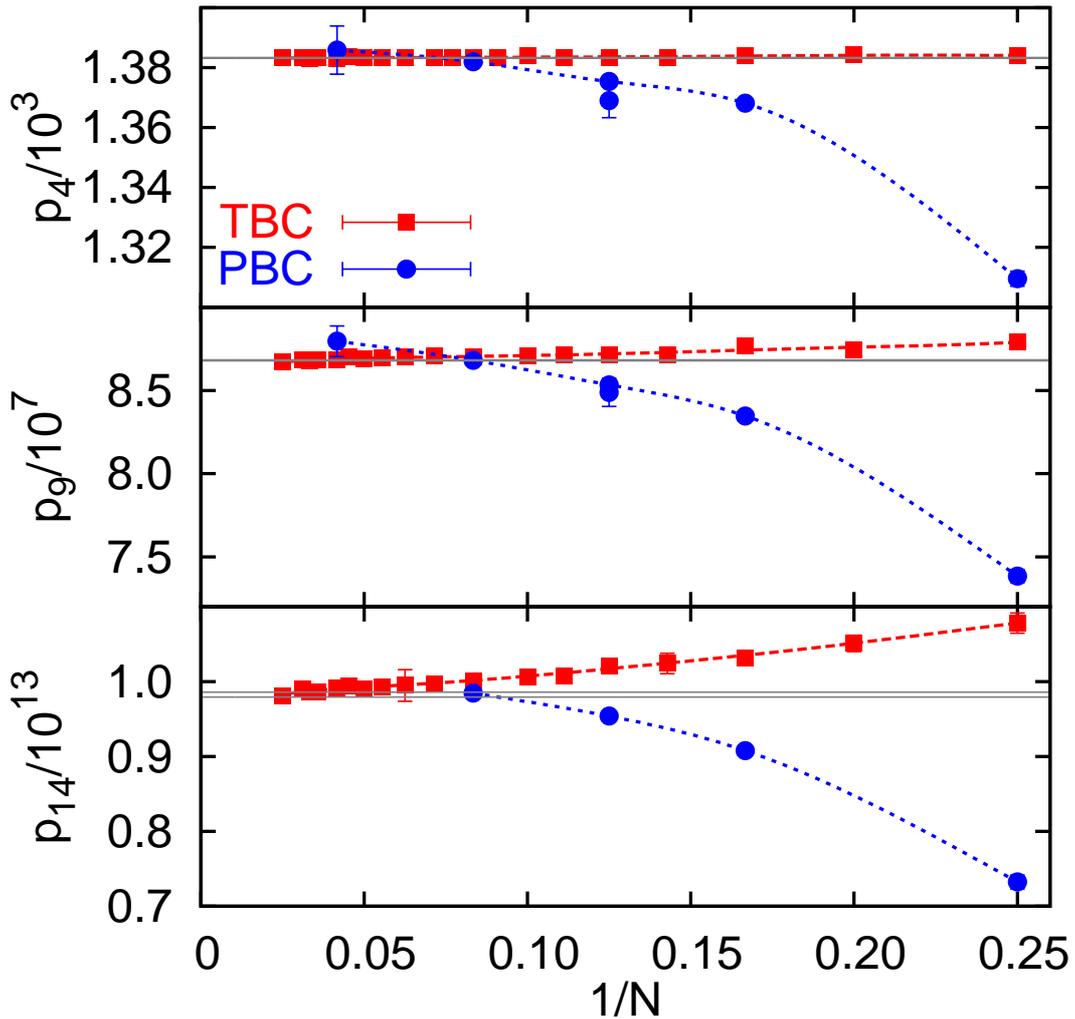}}
\caption{\it \label{fig:PBCvsTBC} 
The coefficients $p_{4,9,14}(N)$ as functions of $1/N$
for the PBC~\protect{\cite{Horsley:2012ra,DiRenzo:2000ua}} (circles) and TBC (squares) data.
The splines are drawn to guide the eye. The grey horizontal error bands
are the infinite volume extrapolated values, see the
last column of Table~\ref{tab:PlaqCoeffserrors}.}
\end{figure}

As detailed in Ref.~\cite{Bali:2013pla}, in addition to the TBC simulations
presented here, for testing purposes and to enable comparison with
literature values, we also performed simulations
employing PBC. These PBC runs however were limited to small
volumes and orders. Therefore, we will
resort to literature values to enable a comparison
between PBC and TBC. In Ref.~\cite{Horsley:2012ra} PBC results
up to $\order(\al^{20})$ were presented for $N=4,6,8,12$.
Up to $\order(\al^{10})$ these can be combined with
earlier $N=8$ and $N=24$ results~\cite{DiRenzo:2000ua}, 
and up to $\order(\al^{3})$  
with $N=32$ results~\cite{DiRenzo:2004ge}.

In Fig.~\ref{fig:PBCvsTBC}, we compare the volume dependence
of the  PBC data from the literature
with our TBC results for the examples of $p_4$, $p_9$ and $p_{14}$.
The horizontal bands denote the
infinite volume extrapolations and their errors, obtained as
will be described in
Sec.~\ref{sec:fits} below and displayed in the last column
of Table~\ref{tab:PlaqCoeffserrors}.
These are independent of the boundary conditions and should be the same,
irrespectively of using PBC or TBC.
The PBC data appear to somewhat over-shoot the infinite volume values.
It is not clear whether this behaviour
can be attributed to a non-monotonous
volume dependence
or to a less well-controlled $\epsilon\rightarrow 0$ extrapolation
of the PBC data, which were obtained using the
unimproved $\order(\epsilon)$ Euler
integration scheme.
It is clear from the comparison that the TBC volume dependence
is much reduced relative to the PBC case. However, at large orders
also the TBC data start to show a significant dependence on $N$.
In the next section, we will discuss
theoretical expectations on the volume dependence
both for TBC and for PBC.

\section{Finite volume corrections}
\label{sec:fse}

In this section we determine the structure of the volume dependence
of the coefficients $p_n(N)$ in the limit of large $N$. 
For simplicity we assume fixed aspect ratios between
different directions, so that the finite volume
effects can only depend on one parameter, $N$. More specifically,
we simulate and consider symmetric lattice volumes.
Together with the symmetry of the action, measure and
observable under the interchange $a \leftrightarrow -a$,
this implies that the coefficients $p_n(N)$ of
\Eqre{eq:expand} are
functions of $N^2$ only:
\be
\label{pnN}
p_n(N)=p_n-\frac{h_n(N)}{N^2}-\frac{f_n(N)}{N^4}-\frac{g_n(N)}{N^6}+\order\left(\frac{1}{N^8}\right)
\,.
\ee

In the following, we will distinguish between TBC and PBC. Below we discuss theoretical expectations
for the two types of boundary conditions, before we
confront the numerical PBC data, where finite volume
effects are more easily detectable, with different parametrizations.

\subsection{Perturbative OPE with TBC}
\label{sec:TBC}
There are no zero modes using TBC (see, for instance
Refs.~\cite{Coste:1985mn,Luscher:1985wf}) and
perturbation theory is characterized by two distinct scales: 
$1/a$ and $1/(Na)\equiv1/\ell$. In this context,
the $N$-dependence of $h_n(N)$, $f_n(N)$ and $g_n(N)$ appears as the ratio of 
these two scales, $a/(Na)$, and 
perturbation theory predicts that it is logarithmic:
\be
\label{fnilog}
h_n(N)=\sum_{i=0}^nh_n^{(i)}\ln^i(N)\,, \quad f_n(N)=\sum_{i=0}^nf_n^{(i)}\ln^i(N)\,,
\quad g_n(N)=\sum_{i=0}^ng_n^{(i)}\ln^i(N)\,.
\ee 
 
We are interested in the large-$N$ (i.e.\ infinite volume)
limit. In this situation 
\be
\label{eq:scales}
\frac{1}{a} \gg \frac{1}{Na}
\ee
and it makes sense to factorize the contributions of the
different scales within the OPE
framework.\footnote{There are rigorous theorems proving the validity of the OPE 
within finite-order perturbation theory
for renormalizable theories~\cite{Zimmermann:1972tv}.}
The hard modes, of scale $\sim 1/a$,
determine the Wilson
coefficients, whereas the soft modes, of scale $\sim 1/\ell$, can be described
by expectation values of local gauge invariant operators.
Due to the absence of such operators of dimension two,
there can be no $1/N^2=a^2/\ell^2$ terms,
i.e.\ $h_n=0$ in \Eqre{pnN}.
The $1/N^4$-term, i.e.\ $f_n(N)$,
is also fixed to a large extent by the OPE. 
The renormalization group invariant definition of the gluon condensate
\be
\label{eq:GC}
\langle O_{\G}\rangle=-\frac{2}{\beta_0}\left\langle\Omega\left| \frac{\beta(\alpha)}{\alpha}
G_{\mu\nu}^cG_{\mu\nu}^c\right|\Omega\right\rangle
=
\left\langle\Omega\left|  \left[1+\order(\alpha)\right]\frac{\alpha}{\pi}
G_{\mu\nu}^cG_{\mu\nu}^c\right|\Omega\right\rangle
\ee
is the only local gauge invariant expectation value of an operator of dimension
$a^{-4}$. In the purely perturbative case discussed here, it only depends on the soft scale $1/\ell$, i.e.\ on the lattice
size. 
On dimensional grounds, the perturbative gluon condensate
$\langle O_{\G}\rangle_{\soft}$ is proportional to 
$1/\ell^4=1/(Na)^4$, and the logarithmic $\ell$-dependence is encoded
in  $\al(\ell^{-1})$. Therefore, 
\be
\label{eq:fnnn}
\frac{\pi^2}{36}\,a^4\langle O_{\G}\rangle_{\soft}=
-\frac{1}{N^4}
\sum_{n\geq 0}f_n\al^{n+1}((Na)^{-1})\,,
\ee
and the perturbative expansion of the plaquette on a finite
volume of $N^4$ sites can be written as\footnote{
On the lattice the continuum O(4)
symmetry is broken down to the hypercubic subgroup H(4).
The corrections due to this however are of size $(a^2/\ell^2)/N^4=1/N^6$
and will only show up in the next order of the OPE.
In particular this means that more than one matrix element of dimension
six needs to be considered.}
\be
\label{OPEpert}
\langle P \rangle_{\pert} (N)=
P_{\pert}(\al)\langle\mathds{1}\rangle
+\frac{\pi^2}{36}C_{\G}(\al)\,a^4\langle O_{\G}\rangle_{\soft}
+\order\left(\frac{1}{N^6}\right)\,,
\ee
where 
\be
P_{\pert}(\al)=\sum_{n\geq 0}p_n\al^{n+1}
\ee
and $p_n$ are the infinite volume coefficients that we
are interested in. The constant prefactor $\pi^2/36$ is chosen
such that the Wilson coefficient, 
which only depends on $\al$, 
is normalized to unity for $\al=0$. It can be expanded as follows:
\begin{align}
C_{\G}(\al)&=1+\sum_{k\geq 0}c_k\al^{k+1}
\,.
\end{align}

Combining the above three equations gives
\begin{align}
\label{eq:Plaq2}
\langle P \rangle_{\pert}(N)
&=\sum_{n\geq 0}\left[p_n-\frac{f_n(N)}{N^4}\right]\al^{n+1}
\\\nn&=
\sum_{n\geq 0}p_n\al^{n+1}
\\\nn
&
\quad
-\frac{1}{N^4}
\left(1+\sum_{k\geq 0}c_k\al^{k+1}(a^{-1})
\right)
\times 
\sum_{n\geq 0}f_n\al^{n+1}((Na)^{-1})
+\order\left(\frac{1}{N^6}\right)\,,
\end{align}
where ultimately we are interested in the $p_n$.
Comparing the above expression with \Eqre{fnilog}, we obtain $f_n^{(i)}$ as polynomials of $f_j$ 
and $c_k$:
\begin{align}
\label{eq:fn0}
f_0(N)&=f_0\,,
\\
\label{f1NV}
f_1(N)&=\left(f_1+c_0f_0\right)+f_0\frac{\beta_0}{2\pi}\ln(N)\,,
\\
f_2(N)&=\left(f_2+c_0f_1+c_1f_0\right)+\left[\left(2f_1+c_0f_0\right)\frac{\beta_0}{2\pi}+f_0\frac{\beta_1}{8\pi^2}\right]
\ln(N)\\\nn
&+f_0\left(\frac{\beta_0}{2\pi}\right)^{\!\!2}\!\ln^2(N)\,,\\
\label{fnN}
f_n(N)
&=
\left(f_n+c_0f_{n-1}+\cdots+c_{n-2}f_1+c_{n-1}f_0\right)
\\
\nn
&+
\left\{\left[nf_{n-1}+(n-1)c_0f_{n-2}+(n-2)c_1f_{n-3}+\cdots+c_{n-2}f_0\right]\frac{\beta_0}{2\pi}+\cdots \right\}
\ln(N)
\\
\nn
&+\cdots\,.
\end{align}
The $\beta$-function coefficients 
and the logarithms above are obtained by expanding
$\al((Na)^{-1})$ within \Eqre{eq:Plaq2} in terms of $\al=\al(a^{-1})$
using the renormalization group, where we define the QCD $\beta$-function as
\be
\label{betafunction}
\beta(\al(\mu))=\frac{d\al(\mu)}{d\ln\mu}=-2\alpha\left[
 \beta_0\frac{\al(\mu)}{4\pi}
+\beta_1\left(\frac{\al(\mu)}{4\pi}\right)^2 +\cdots\right]\ ,
\ee
where
\begin{align}
\beta_0&=11\,,\quad\beta_1=102\,,\nonumber\\\label{eq:betas}
\quad\beta_2^{\MS}&=\frac{2857}{2}\,,\quad\beta_2^{\latt}=-6299.8999(6)\,,\\
\beta_3^{\MS}&\approx 29243.0\,,\quad \beta_3^{\latt}= -1.16(12)\times 10^6\,.
\nonumber\end{align}
$\beta_3^{\MS}$ was calculated in
Ref.~\cite{vanRitbergen:1997va} where the previous results
on $\beta_0$, $\beta_1$ and $\beta_2^{\MS}$ are
referenced.
In the lattice scheme only $\beta_2^{\latt}$ has been
computed diagrammatically~\cite{Luscher:1995np,Christou:1998ws,Bode:2001uz}.
The value for $\beta_3^{\latt}$ that we quote~\cite{Bali:2013qla} was obtained
by calculating the
normalization of the heavy quark pole mass renormalon and
then assuming the corresponding $\MS$-scheme 
expansion to follow its asymptotic behaviour
from orders $\als^4$ onwards. 
Similar estimates, $\beta_3^{\latt}\approx
-1.37\times 10^6$ up to $\beta_3^{\latt}\approx
-1.55\times 10^6$, were found in Ref.~\cite{Guagnelli:2002ia}
using a very different method.

Note that the coefficients $f_n^{(i)}$ within \Eqre{fnilog} for $i>0$
are entirely determined by
$f_j$ and $\beta_j$ with $j<n$
and $c_k$ with $k<n-1$. Eqs.~\eqref{eq:Plaq2} -- \eqref{fnN}
are the most general parametrization
of the $1/N^4$-effects for any
lattice action using TBC. 

Using the above conventions, the trace anomaly of the
energy-momentum tensor reads
\be
\label{eq:enmom}
\langle \Omega|T_{\mu\mu}(0)|\Omega\rangle=
\left\langle\Omega\left| \frac{\beta(\alpha)}{4\alpha}G_{\mu\nu}^c(0)G_{\mu\nu}^c(0)\right|\Omega\right\rangle
=-\frac{\beta_0}{8}\langle O_{\G}\rangle\,,
\ee
which in turn equals the expectation
value of the Lagrangian density times $\beta(\alpha)/\alpha$.
In this paper we employ the Wilson action, for which the
discretized Lagrangian is exactly proportional to the
plaquette $P$, see \Eqre{eq:plaqnorm}, so that the
above relation --- in this case between the
plaquette and $a^4\langle O_{\G}\rangle$
--- holds up to $\order(a^6)$ corrections.
This fixes the Wilson coefficient exactly~\cite{DiGiacomo:1990gy,DiGiacomo:1989id}:
\begin{align}
C_{\G}(\al)&=1+\sum_{k\geq 0}c_k\al^{k+1}=
\label{CP}
-\frac{\beta_0\al^2}{2\pi\beta(\al)}
\\
\nn
&=1-\frac{\beta_1}{\beta_0}\frac{\al}{4\pi}
+\frac{\beta_1^2-\beta_0\beta_2}{\beta_0^2}\left(\frac{\al}{4\pi}\right)^2
-\frac{\beta_1^3-2\beta_0\beta_1\beta_2+\beta_0^2\beta_3}{\beta_0^3}\left(\frac{\al}{4\pi}\right)^3
+\order(\al^4)
\,.
\end{align}
Note that $C_{\G}(\al)$ is
scheme-dependent not only through $\alpha$, but also
explicitly, due to its dependence on the higher $\beta$-function
coefficients $\beta_2$ etc..
The $c_k$ 
depend on the $\beta_i$ with $i\leq k+1$ via \Eqre{CP}.

Finally, we consider $1/N^6$-effects. At this order the number of terms
and thus fit parameters grows quite rapidly.
Therefore, we do not attempt a complete study of
the $1/N^6$ corrections, but aim at achieving a qualitative understanding
of the corresponding structure. The philosophy is the same as above: we have to
carry out the OPE program to the next order. This means that we have
to consider all gauge invariant local operators of dimension six
that are singlets under the hypercubic subgroup
H(4) of O(4).\footnote{The matrix elements depend only on momentum
scales much smaller than 
$1/a$. This is the reason we can use continuum notations for
the matrix elements. The physics associated to the scale $1/a$
is encoded in the Wilson coefficients.} 
Three such operators exist~\cite{Luscher:1984xn},
one of which can be eliminated via the equations of
motion for on-shell quantities.
We consider the O(4) invariant
$\langle O_6\rangle=\langle g G^3 \rangle$ as one
such example but in principle a
second matrix element needs to be added.
$O_6$ has a non-trivial anomalous dimension, 
complicating the logarithmic corrections.
The contribution of this term will be 
\begin{align}\nn
\delta \langle P \rangle_{\pert}(N) &
\sim \frac{1}{N^6}
\left[1+\sum_{k\geq 0}c^{(6)}_k\al^{k+1}(a^{-1})\right]
\times 
 \exp\left[\int_{\al(a^{-1})}^{\al((Na)^{-1})}\frac{d\al^\prime}{\al^\prime}\left(\gamma_0+\gamma_1\al^\prime+\cdots\right)\right]
\\\label{FSE6}&\quad\times
\sum_{n\geq 0}g_n\al^{n+1}((Na)^{-1})\,.
\end{align}
$\gamma_0$, the one-loop anomalous dimension of $O_6$, is known~\cite{Narison:1983kn} 
but not the higher orders in the scheme we use. The above structure results in
three new unknown parameters for each additional power of $\al$:
one additional $c_k^{(6)}$-value for the Wilson coefficient,
one higher order anomalous dimension coefficient $\gamma_j$
and an additional $g_n$-value from the expansion of
$\langle O_6\rangle_{\soft}$.

Besides the OPE of the plaquette expectation value, we also have to
perform the OPE of the lattice action, to obtain an effective action 
where only soft modes remain dynamical:
\be
S=\frac{1}{4}\int\!d^4\!x\,G^2(x)+a^2C_6(a^{-1})\int\! d^4\!x\, gG^3(x)+\cdots\,.
\ee 
The dimension six operators here are the same as those considered above, since the symmetries are the same. 
Again we focus on $O_6$, which produces
the following additional contribution
to $P_{\pert}(N)$:
\begin{align}
\tilde{\delta} \langle P \rangle_{\pert}(N) &\sim a^6C_6(a^{-1})\,\int\! d^4\!y\,\langle \mathcal{T}\!\left\{G^2(0),O_6(y)\right\}\rangle_{\soft}
\\
\nn
&\sim
\frac{1}{N^6}
\left[1+\sum_{k\geq 0}\tilde{c}^{(6)}_k\al^{k+1}(a^{-1})
\right]
\times 
 \exp\left[\int_{\al}^{\al((Na)^{-1})}\frac{d\al^\prime}{\al^\prime}\left(\gamma_0+\gamma_1\al^\prime+\cdots\right)\right]\\\nn
& \times
\sum_{n\geq 0}\tilde{g}_n\al^{n+1}((Na)^{-1})\,.
\end{align}
The anomalous dimension is the same as that in \Eqre{FSE6},
as the operator is the same. Since we
employ the plaquette action, also the Wilson coefficient is 
identical to that in \Eqre{FSE6} ($\tilde{c}^{(6)}_k={c}^{(6)}_k$) and
differences between the soft matrix elements
can be absorbed into \Eqre{FSE6}, redefining $g_n+\tilde{g}_n\mapsto
g_n$.
Therefore, no additional parameters are required. The same arguments also
apply to the second independent operator of dimension six.\footnote{
Note that this second dimension six operator is not invariant under O(4) spacetime rotations~\cite{Luscher:1984xn}.}
Overall, at $\order(1/N^6)$ we expect a total of six new parameters per order
in $\al$, which exceeds our fitting capabilities.  
Therefore, we do not attempt a more systematic study of the $1/N^6$-effects. 

\subsection{Non-perturbative OPE with TBC}

Since in NSPT we Taylor expand in powers of $g$ before averaging
over the gauge variables, no  
mass gap is generated dynamically.
It is interesting though to discuss in what
particular setting our results can be related to
non-perturbative results obtained by
Monte-Carlo lattice simulations. 
In this case an additional 
scale, $\lQ \sim 1/a \, e^{-2\pi/(\beta_0\al)}$,
is generated dynamically.
However, we can always tune $N$ and $\al$ such that
\be
\label{eq:scalesNP1}
\frac{1}{a} \gg \frac{1}{Na}=\frac{1}{\ell}  \gg \lQ\,.
\ee
In this small-volume situation we encounter a double expansion in powers of 
$a/\ell$ and $a\lQ$ (or $(\ell\lQ)(a/\ell)$). The construction of the OPE
is completely analogous to that of Sec.~\ref{sec:TBC} above
and we obtain\footnote{
In the last equality, we approximate the Wilson coefficients by their
perturbative expansions,
neglecting the possibility of non-perturbative contributions
associated to the hard scale $1/a$. These  
would be suppressed by factors $\sim\exp(-2\pi/\alpha)$
and therefore would be subleading relative to the gluon condensate.}
\begin{align}
\label{OPEMC}
\langle P\rangle_{\MC} &=\frac{1}{Z}\left.\int\![dU_{x,\mu}]\,e^{-S[U]} P[U]
\right|_{\MC}
\\\nn
&=
P_{\pert}(\al)\langle\mathds{1}\rangle
+\frac{\pi^2}{36}C_{\G}(\al)\,a^4\langle O_{\G}\rangle_{\MC}
+\order\left(a^6\right)\,.
\end{align}
In the last equality we have factored out the hard scale, $1/a$, from the 
scales $1/(Na)$ and $\lQ$, which are encoded in
$\langle O_{\G}\rangle_{\MC}$. Exploiting the right-most
inequality of \Eqre{eq:scalesNP1},
we can expand $\langle O_{\G}\rangle_{\MC}$ as follows:
\be
\langle O_{\G}\rangle_{\MC}=\langle O_{\G}\rangle_{\soft}\left[1+\order(\lQ^2 \ell^2)\right]
\,.
\ee
Hence, a non-perturbative
small-volume simulation\footnote{Also in this case one encounters
technical problems that are resolved
using TBC, see Ref.~\cite{Trottier:2001vj}.}
would yield the same expression as NSPT,
up to non-perturbative corrections that can be made arbitrarily
small by reducing $a$ and therefore
$\ell=Na$, keeping $N$ fixed.
In other words, $p_n^{\NSPT}(N)=p_n^{\MC}(N)$ up to
non-perturbative corrections.

We can also consider the limit 
\be
\label{eq:scalesNP2}
\frac{1}{a}  \gg \lQ \gg \frac{1}{Na} \,.
\ee
This is the standard situation realized in non-perturbative lattice simulations.
Again the OPE can be constructed as in Sec.~\ref{sec:TBC} and \Eqre{OPEMC}
also holds. The difference is that now 
\be
\langle O_{\G}\rangle_{\MC}=
\langle O_{\G}\rangle_{\NP}\left[1+\order\left(\frac{1}{\lQ^2 \ell^2}\right)
\right]
\,,
\ee
where $\langle O_{\G}\rangle_{\NP} \sim \lQ^4$ is the so-called
non-perturbative gluon condensate introduced
in Ref.~\cite{Vainshtein:1978wd}.

Finally, we re-emphasize that \Eqre{OPEMC} holds, irrespectively of 
ordering the scales according to \Eqre{eq:scalesNP1} or to \Eqre{eq:scalesNP2}.
We further remark that the 
relation \Eqre{CP}
for the Wilson coefficient $C_{\G}$ for the plaquette action
also holds when non-perturbative effects
are included.

\subsection{Perturbative and non-perturbative OPEs with PBC}
\label{sec:PBC}

In the case of PBC one encounters constant, i.e.\ zero, modes.
The effects associated to these are non-perturbative in nature.
They can be interpreted as introducing an extra scale $g^{1/2}/(Na)$, 
besides the perturbative scales $1/a$ and $1/(Na)$.
Therefore, with PBC, irrespectively of how small the coupling is,
there are non-perturbative effects associated with these
modes,\footnote{As with TBC, we could also admit $\lQ$ into our considerations
as long as the hierarchy \Eqre{eq:hira}
is satisfied.} which will invalidate the perturbative OPE
of the plaquette with PBC. 
The violations of the perturbative OPE will decrease with $1/N^4$ because
the relative measure of the zero mode contributions becomes
suppressed by this factor for large volumes.
These effects are then of the same order as those associated
with $\langle O_{\G}\rangle_{\soft}$.
Both contributions will undergo mixing
and invalidate the parametrization of the finite size effects
Eqs.~\eqref{eq:Plaq2} -- \eqref{fnN}.

The $\order(\al)$
zero mode contribution has been explicitly computed
in Ref.~\cite{Coste:1985mn}. Generalizing this derivation
to higher orders in $\al$ becomes extremely complicated.
In particular one has to disentangle the contributions of
the different scales. Since it is not clear how to
properly account for the zero modes, in practice
they are omitted in diagrammatic PBC calculations or
subtracted in NSPT computations. In particular, the literature
results of $p_n^{\PBC}(N)$ that we use here do not include
these contributions. Therefore, these literature values
do not correspond to any physical situation,
except in the infinite
volume limit where zero modes can be neglected.
In other words, the coefficients $p_n^{\PBC}(N)$
cannot be obtained from a fit to non-perturbative data
(with infinite precision) of the plaquette computed in the situation  
\be
\label{eq:hira}
\frac{1}{a} \gg \frac{1}{Na} \gg \frac{g^{1/2}}{Na} \gg \lQ\,.
\ee
This means that one cannot apply the standard OPE
and the finite size behaviour of the $p_n^{\PBC}(N)$ is less well
constrained than in the TBC case.
However, the leading-order corrections will still scale as $1/N^4$, 
and they will be logarithmically modulated.
Given precise data and large volumes, this may still suffice
to extrapolate high-order coefficients $p_n(N)$ to infinite $N$.

\subsection{Phenomenological fits to PBC data}
\label{sec:fitPBC}

In order to confirm the validity of the interpolating function
and the perturbative OPE structure discussed above,
we perform a series of tests using the PBC data. In particular we investigate
numerically whether any $1/N^2$ effects, which are
incompatible with the expected OPE structure, may nevertheless be present
in the data or in diagrammatic lattice perturbation theory.

We start by studying the low-order coefficients obtained
using diagrammatic lattice perturbation theory.
At $\order(\al)$ exact results can be derived both for PBC
and for TBC:\footnote{We remark again that the
PBC result is obtained omitting the zero mode contribution.}
\be
\label{p0NV}
p^{\TBC}_0(N)=\frac{4}{3}\pi 
\,,
\qquad p^{\PBC}_0(N)=\frac{4}{3}\pi\left(1-\frac{1}{N^4}\right)
\,.
\ee
One consequence of using TBC instead of PBC is that the one-loop
behaviour is flat: 
$p_0^\TBC\left(N\right)=p\left(\infty\right)\equiv p_0$. 
In Fig.~\ref{fig:DATA_p0} we compared our TBC 
$p_0\left(N\right)$ data with the analytical value and
found agreement within errors down to the smallest lattice volume, so 
finite volume effects are truly absent at leading order. 

The $\order(\al^2)$ infinite volume coefficient was first computed in
Ref.~\cite{DiGiacomo:1981wt} and with increased precision
in Ref.~\cite{Alles:1998is}.
We have recomputed it using the formulae of this last reference
together with the very precise lattice integrals of Ref.~\cite{Luscher:1995np}, 
obtaining
\be
\label{p1infty}
p_1=5.355009398(6)
\,.
\ee
In order to study the $N$-dependence we have also computed
$p^{\PBC}_1(N)$ for $N\leq 64$ and high precision,
using the formulae given in Ref.~\cite{Heller:1984hx}. 
From this analysis we conclude that to this order
there are no $1/N^2$ effects and we obtain
\be
\label{p1NV}
p^{\PBC}_1(N) \approx p_1-\frac{1}{N^4}\left[3.3\ln(N)+13.4\right]-\frac{18}{N^6}\,,
\ee
where we have fixed the $p_1$-value to \Eqre{p1infty}.

Comparing
Eqs.~\eqref{p1NV} and \eqref{p0NV} with \Eqre{f1NV},
we observe that the coefficient of $\ln(N)$ does not comply with the OPE
($3.3 \neq\beta_0f_0=22/3$). This difference illustrates
that we cannot use the OPE with PBC after subtracting the zero modes.
The zero modes contribute to the $\order(\al)$ constant
as well as to the logarithmic and constant terms at
$\order(\al^2)$ (at higher orders the contribution
could be more complicated, due to the $g^{1/2}/(Na)$ scale):
\be
\delta p^{\mathrm{zero\; mode}}_1(N)= \frac{1}{N^4}\left[a\ln(N)+b\right]
+\order\left(\frac{1}{N^6}\right)\,.
\ee
This term was partially subtracted by omitting zero momentum
contributions to the lattice sums. In any case, at present nothing about
the coefficients $a$ or $b$ is known.
Based on this diagrammatic perturbation theory analysis for PBC
we conclude that there are no $1/N^2$ effects at $\order(\al)$ nor
at $\order(\al^2)$. We remark that there are
indications\footnote{We thank H.~Panagopoulos for this comment.} that these may also be absent at $\order(\al^3)$, for which the infinite volume coefficient
was first computed in Ref.~\cite{Alles:1993dn} and with increased
precision in Ref.~\cite{Alles:1998is}:
 \be
 p_2 = 27.1983(9)\,.
 \ee

We now turn to the NSPT PBC data. These cover orders up to
$\al^{20}$. We have seen in Sec.~\ref{sec:quali} (see Fig.~\ref{fig:PBCvsTBC})
that the dependence on $1/N$ is much more pronounced with PBC than with TBC. 
While this additionally complicates the infinite volume
extrapolation of PBC results, it allows us to identify the power
scaling of the leading $1/N$ correction with higher numerical
significance than for TBC.

\begin{table}
\caption{\it Exploratory fits to PBC data
and the resulting $\chired$ as a measure of the fit quality.
All fits have two parameters per order $n$.
The finite size correction is $f_n(N)/N^d$.
The second column is for $f_n(N)$ with renormalization group
running while the third column is for constant $f_n(N)=f_n$.\\~}
\label{tab:DATAPBC_chi2}
\begin{ruledtabular}
\begin{tabular}{>{$}c<{$}|>{$}c<{$}>{$}c<{$}}
\mathrm{power}\,\, d~~ &\chired\,\,(\mathrm{run})&\chired\,\,(\mathrm{no\,\,run})\\\hline
2 & 63.40& 20.19\\
4 & 4.24 &  7.45\\
6 & 11.01 & 22.79\\
\end{tabular}
\end{ruledtabular}
\end{table}

We attempt several fits to PBC $N\geq 4$ data,
assuming the leading term to be of the form
$p_n-f_n(N)/N^d$ with $d=2,4,6$, where we allow for two
different parametrization of $f_n(N)$: $f_n(N)=\mbox{const.}=f_n$ (no run),
and $f_n(N)$ as given in Eqs.~\eqref{eq:fn0} -- \eqref{fnN} (run),
setting $c_n=0$. In each of these parametrizations
we encounter two fit parameters, $p_n$ and $f_n$, per order
of the expansion. The resulting reduced $\chi^2$-values
$\chired\equiv \chi^2/N_{\rm DF}$ (as a measure of
the quality of the respective fits) are shown in the second
and third columns of Table~\ref{tab:DATAPBC_chi2}. 
The numbers indicate that the parametrizations work best for $d\sim4$.
Higher and, most notably, lower values of $d$ 
are clearly ruled out by the data, irrespectively of 
including a running into the $f_n(N)$ or not.
We also see that for $d=4$ the data prefer ``running'' to
``no running''.\footnote{
The necessity of a logarithm was also clearly established in the
diagrammatic $p_1(N)$ result \Eqre{p1NV}.} However, we have neglected
the Wilson coefficient of the gluon condensate (the $c_k$),
ignored the (unknown) effect of the subtracted zero modes
and most of the literature data are available only
for rather small $N$ ($=4, 6, 8$ and 12).
Therefore, it is not surprising that
the value $\chired\approx 4.2$ in the best ``running''
$d=4$ case is still unsatisfactory. The number of parameters needed to 
incorporate these effects into the parametrization
will quickly explode with the order, turning a
model-independent fit to PBC data
impossible for any realistic number of volumes.

We conclude that no $1/N^2$ terms exist and that
some sort of running of the $1/N^4$-term is required
to describe the PBC data. We take this as a confirmation of the theoretical
arguments presented in Sec.~\ref{sec:TBC}.

\section{Infinite volume coefficients}
\label{sec:fits}
In this section we determine the infinite volume coefficients $p_n$,
defined in \Eqre{pnN}, for $0<n \leq 34$. For $n=0$, we
use the exact result $p_0=4\pi/3$.
Our default fit function for $p_n(N)$ is (see also \Eqre{eq:Plaq2})
\be
\label{defaultfit}
p_n(N)=p_n-\frac{f_n(N)}{N^4}
\,,
\ee
where the $f_n(N)$ are defined
in Eqs.~\eqref{eq:fn0} -- \eqref{fnN}.
$p_n(N)$ depends on the fit parameters
$p_n$, $f_j$ with $j\leq n$, and $c_k$, with $k\leq n-1$.
We know from diagrammatic calculations that $f_0=0$.
Since $f_0=0$, $c_{33}$ does not appear in the fit. We will also
set $c_{32}=0$, as this coefficient cannot be parametrically
distinguished from $f_{34}$. 
For the $\beta$-function coefficients that appear in our fit function,
we will set 
$\beta_0$, $\beta_1$ and $\beta_2$ 
to their known values \Eqre{eq:betas}
(note that $\beta_2$ depends on the scheme)
and $\beta_i=0$ for $i\geq 3$. We also fix 
$c_0$ and $c_1$ to their known values of \Eqre{CP}
($c_1$ is scheme-dependent too).
Therefore, our default fit function depends on 
a total of 34 $p_n$-coefficients,
34 $f_n$-coefficients, and 30 $c_n$-coefficients.
This function with 98 free parameters should
describe all 35 orders of perturbation theory on the volumes
listed in Table~\ref{tab:volumesPlaq} for any $N$ bigger
than a small volume cut-off $\nu\leq N$. 15 different volumes will contribute to
our primary fit, described below.

The combined dependence 
on $f_j$ and $c_k$ introduces strong correlations between different
orders, which we take into account by simultaneously fitting
all $p_n(N)$ for $0<n \leq 34$.
Unlike in Ref.~\cite{Bali:2013pla}, we
cannot, in a first sweep, fit each new $p_n(N)$ independently with
two new fit parameters $f_n$ and $c_{n-1}$, keeping the
$f_j$- and $c_k$-values that were obtained at previous orders $k<j<n$ fixed
and, subsequently, run the fit to convergence. The reason is that the $c_k$
non-linearly couple different orders, which
considerably complicates the fitting procedure. Particularly
problematic is the introduction of 
the $c_k$ for small values of $k$, which makes finding stable solutions quite difficult (with a large region of the parameter space of $c_k$ and $f_j$
producing small variations of $\chired$). 
This is so
because the parametrization cannot easily distinguish between, for instance,
$c_0f_{33}\al(a^{-1})\al^{34}((Na)^{-1})$ and 
$f_{34}\al^{35}((Na)^{-1})$, as the running of these two terms is
very similar. This problem is alleviated because we
know $c_0$ and $c_1$ analytically. Fortunately, as we increase the order
$k$ of $c_k$
the running of different products
$c_kf_{n-k}\al^n(a^{-1})\al^{n-k+1}((Na)^{-1})$ becomes
more and more distinguishable.

\begin{table}[hb]
\ra{0.73}
\caption{\it $\chired$ and $p_n$ for different
values of $\nu$ using the fit function \Eqre{defaultfit}. The $n=0$ values
were fixed to the exact result. The diagrammatic expectations are 
$p_1=5.355009398(6)$ and $p_2=27.1983(9)$.\ }
\label{tab:PlaqCoeffsVol}
\begin{ruledtabular}
\begin{tabular}{>{$}c<{$}|>{$}c<{$}>{$}c<{$}>{$}c<{$}>{$}c<{$}}
\nu & 13& 11&9 &7\\\hline
\chired    &0.826107         & 0.768641         &  0.700803 & 0.863024\\
p_0 & \mathbf{4{\boldsymbol\pi}/3}           &\mathbf{4{\boldsymbol\pi}/3} & \mathbf{4{\boldsymbol\pi}/3}&\mathbf{4{\boldsymbol\pi}/3}\\
p_1& 5.35606(66)                & 5.35539(25)   & 5.35522(13)    & 5.35509(12)\\
p_2/10& 2.71947(22)          &2.71978(14)    & 2.719761 (94) & 2.719752(81)\\
p_3/10^{2}  & 1.80963(13)   & 1.809690(92) & 1.809718(73)  & 1.809747(64) \\
p_4/10^{3}  & 1.38319(23)   & 1.38324(14)   & 1.383242(90)    &1.383285(75)\\
p_5/10^{4}  & 1.15170(53)   & 1.15189(42)   & 1.15184(26)    &1.15186(12) \\
p_6/10^{5}  & 1.01632(63)   & 1.01650(52)   & 1.01678(35)    &1.01670(16)\\
p_7/10^{5}  & 9.3553(64)     & 9.3572(54)     & 9.3605(40)    &9.3621(21) \\
p_8/10^{6}  & 8.8936(53)     & 8.8949(44)     & 8.8971(35)      &8.9025(23)\\
p_9/10^{7} & 8.6745(43)      & 8.6752(39)     & 8.6803(29)      &8.6858(27)\\
p_{10}/10^{8} & 8.6358(74)  & 8.6370(67)    &8.6441(53)       &8.6532(45)\\
p_{11}/10^{9}  & 8.744(14)   & 8.745(12)      &8.7568(97)         &8.7706(74)\\
p_{12}/10^{10} & 8.977(25)  & 8.979(20)      & 8.993(16)        &9.003(14) \\
p_{13}/10^{11}  & 9.331(38) & 9.331(30)      & 9.350(23)        &9.366(19) \\
p_{14}/10^{12} & 9.805(54)  & 9.796(43)      & 9.827(33)        &9.847(28)\\
p_{15}/10^{14} & 1.0397(78)& 1.0382(63)    & 1.0423(46)      &1.0448(39)\\
p_{16}/10^{15} & 1.111(12)  & 1.110(10)    & 1.1143(69)        &1.1173(57)\\
p_{17}/10^{16} & 1.196(19)  & 1.194(16)      & 1.201(10)        &1.2041(84)\\
p_{18}/10^{17} & 1.294(29)  & 1.294(26)      & 1.303(15)        &1.307(12) \\
p_{19}/10^{18} & 1.409(44)  & 1.416(39)      & 1.421(22)        &1.426(18)\\
p_{20}/10^{19} & 1.544(64)  & 1.554(57)      & 1.562(32)        &1.567(25)\\
p_{21}/10^{20} & 1.704(93)  & 1.723(82)      & 1.727(44)        &1.731(35)\\
p_{22}/10^{21}  & 1.89(13)   & 1.93(12)      & 1.924(61)          &1.922(48)\\
p_{23}/10^{22} & 2.11(19)    & 2.20(16)        & 2.160(84)          &2.143(69)\\
p_{24}/10^{23} & 2.38(28)    & 2.54(23)        & 2.45(12)          &2.40(10)  \\
p_{25}/10^{24}  & 2.76(40)   & 3.02(33)        & 2.82(18)          &2.71(15)\\
p_{26}/10^{25} & 3.31(58)    & 3.71(50)        & 3.32(28)          &3.10(24)\\
p_{27}/10^{26} & 4.14(85)    & 4.79(87)        & 4.04(46)          &3.60(40)\\
p_{28}/10^{27} & 5.4(13)      & 6.6(17)          & 5.15(82)            &4.32(67) \\
p_{29}/10^{28} & 7.6(22)      & 9.6(33)          & 7.0(15)            &5.5(11) \\
p_{30}/10^{30} & 1.20(43)    & 1.55(67)        & 1.04(29)          &0.76(21) \\
p_{31}/10^{31} & 1.90(84)    & 2.5(13)        & 1.64(55)          &1.15(38) \\
p_{32}/10^{32} & 3.1(17)      & 4.3(26)          & 2.7(11)            &1.90(70) \\
p_{33}/10^{33} & 5.2(33)      & 7.3(52)          & 4.8(20)            &3.3(14)\\
p_{34}/10^{34} & 9.1(65)      & 13(10)          & 8.8(40)            &6.4(27) \\
\end{tabular}
\end{ruledtabular}
\end{table}

\begin{table}[hb]
\ra{0.73}
\caption{\it The same as Table~\protect\ref{tab:PlaqCoeffsVol}. $\nu=9$, except
for the first column: adding the $1/N^6$-term (\Eqre{N6term}) and
setting $\nu=8$. Second column:
including $\beta_3^{\latt}$ of \Eqre{eq:betas}.
Third column: setting $\beta_2^{\latt}=0$.
Last column: result of Table~\protect\ref{tab:PlaqCoeffsVol},
including systematic errors.}
\label{tab:PlaqCoeffserrors}
\begin{ruledtabular}
\begin{tabular}{>{$}c<{$}|>{$}c<{$}>{$}c<{$}>{$}c<{$}>{$}c<{$}}
                                             & 1/N^6    & \beta_3\neq 0           &\beta_2=0         & \mathrm{Final\; result}\\
                                             \hline
\chired    & 0.671138          &  0.70238       & 0.70026             &\\
p_0                                      & \mathbf{4{\boldsymbol\pi}/3}  & \mathbf{4{\boldsymbol\pi}/3}&\mathbf{4{\boldsymbol\pi}/3}    &\mathbf{4{\boldsymbol\pi}/3}\\
p_1                                      & 5.35559(23)      & 5.35522(13)    & 5.35522(13)       &5.35522(18)\\
p_2/10                                 & 2.71974(14)      & 2.719762(94)  & 2.719761(94)     &2.719761(95)\\
p_3/10^{2}                           & 1.80966(10)      & 1.809719(73)  & 1.809718(73)     &1.809718(78)  \\
p_4/10^{3}                           & 1.38317(15)      & 1.383248(90)  & 1.383244(90)     &1.38324(10) \\
p_5/10^{4}                           & 1.15152(45)      & 1.15164(11)    & 1.15184(26)       &1.15184(26)\\
p_6/10^{5}                           & 1.01617(55)      & 1.01694(32)    & 1.01677(35)        &1.01678(36)\\
p_7/10^{5}                           & 9.3553(55)        & 9.3620(38)      & 9.3604(40)          &9.3605(43)\\
p_8/10^{6}                           & 8.8924(44)        & 8.8978(34)      & 8.8970(35)          &8.8971(65)\\
p_9/10^{7}                           & 8.6729(34)        & 8.6800(27)      & 8.6803(29)          &8.6803(62)\\
p_{10}/10^{8}                       & 8.6331(61)       & 8.6425(50)      & 8.6440(53)           &8.644(11)\\
p_{11}/10^{9}                       & 8.741(11)         & 8.759(10)        & 8.7565(97)           &8.757(17)\\
p_{12}/10^{10}                     & 8.980(19)         & 8.998(15)        & 8.992(16)             &8.993(19)\\
p_{13}/10^{11}                     & 9.339(30)         & 9.355(22)        & 9.350(23)             &9.350(28)\\
p_{14}/10^{12}                     & 9.819(45)         & 9.833(31)        & 9.827(33)             &9.827(38)\\
p_{15}/10^{14}                     & 1.0424(65)       & 1.0427(45)      & 1.0422(47)           &1.0423(53)\\
p_{16}/10^{15}                     & 1.1162(95)       & 1.1150(64)      & 1.1143(69)           &1.1143(75)\\
p_{17}/10^{16}                     & 1.204(14)         & 1.2024(91)      & 1.201(10)             &1.201(11)\\
p_{18}/10^{17}                     & 1.309(20)         & 1.305(13)        & 1.303(15)             &1.303(16)\\
p_{19}/10^{18}                     & 1.433(28)         & 1.424(20)        & 1.421(22)             &1.421(23)\\
p_{20}/10^{19}                     & 1.579(39)         & 1.565(28)        & 1.562(32)             &1.562(32)\\
p_{21}/10^{20}                     & 1.745(56)         & 1.727(41)        & 1.727(45)             &1.727(44)\\
p_{22}/10^{21}                     & 1.955(81)         & 1.921(56)        & 1.924(62)             &1.924(61)\\
p_{23}/10^{22}                     & 2.21(12)           & 2.155(77)        & 2.160(86)             &2.160(86)\\
p_{24}/10^{23}                     & 2.55(19)           & 2.44(11)          & 2.45(12)               &2.45(13)\\
p_{25}/10^{24}                     & 3.02(31)           & 2.80(16)          & 2.83(18)               &2.82(21)\\
p_{26}/10^{25}                     & 3.71(50)           & 3.26(23)          & 3.33(28)               &3.32(35)\\
p_{27}/10^{26}                     & 4.78(84)           & 3.92(37)          & 4.06(47)               &4.04(63)\\
p_{28}/10^{27}                     & 6.6(15)             & 4.92(63)          & 5.19(83)               &5.15(12)\\
p_{29}/10^{28}                     & 9.7(26)             & 6.6(11)            & 7.1(15)                 &7.0(22)\\
p_{30}/10^{30}                     & 1.57(47)           & 9.6(21)            & 1.05(29)               &1.04(40)\\
p_{31}/10^{31}                     & 2.60(87)           & 1.49(40)          & 1.66(56)               &1.64(74)\\
p_{32}/10^{32}                     & 4.4(16)             & 2.46(76)          & 2.8(11)                 &2.7(13)\\
p_{33}/10^{33}                     & 7.6(30)             & 4.2(15)            & 4.9(21)                 &4.8(25)\\
p_{34}/10^{34}                     & 13.6(55)           & 7.7(29)            & 9.0(41)                 &8.8(46)\\
\end{tabular}
\end{ruledtabular}
\end{table}

Using the setup described above, we fit to subsets of data constrained
by $N \geq \nu$, and vary $\nu$. We display some of
these results in Table~\ref{tab:PlaqCoeffsVol} and
use them to explore the validity range of \Eqre{defaultfit}.
Our ``thermometer''
for this will be to obtain acceptable $\chired$-values and
agreement with $p_1$ and $p_2$ from diagrammatic lattice perturbation theory.
We find that including small volumes improves the quality of the
fit down to a cut-off $\nu=9$.
For smaller values of $\nu$ the  $\chired$-values rapidly increase.
This we interpret as becoming
sensitive to higher order finite volume effects that are not accounted
for in our parametrization.
Therefore, we take the results from the $\nu=9$ fit, which uses 365 data points, as our central
values.\footnote{We attribute the fact that $\chired<1$ to our possibly
over-conservative
error estimation for the $p_n(N)$ data.} 

We now estimate the systematic\footnote{
This means, systematic uncertainties other than those of the finite
Langevin step size, discussed in Sec.~\ref{sec:langevin} above,
which are already included
into our ``statistical'' errors.} errors. They are due to 
our incomplete parametrization of
the finite volume corrections, since we have set 
higher $\beta$-function coefficients to zero within the
$1/N^4$ terms. Moreover, 
we have ignored $1/N^6$- and higher order
finite volume corrections.

We determine the $\order(1/N^4)$ truncation
uncertainties in two ways. First we consider the differences
between the central values of the $\nu=9$ and 
$\nu=7$ fits shown in Table~\ref{tab:PlaqCoeffsVol}.
The other possibility we explore is varying the
parametrization to check the robustness of our results.
In principle, the leading parametric uncertainty originates from 
the omission of the higher
order $\beta$-function coefficients: $\beta_3$, $\beta_4$, etc., 
which affect the log-structure of the $1/N^4$ corrections.
Therefore, we perform alternative fits either eliminating
$\beta_2$ (we also set $\beta_2=0$ in $c_1$)
or incorporating $\beta_3^{\latt}$ (quoted in \Eqre{eq:betas})
into our fits. For the first case the outcome is given in the third column of
Table~\ref{tab:PlaqCoeffserrors}. We observe that
the shifts are much smaller than the statistical errors
or the differences between the $\nu=9$ and $\nu=7$ results.
Including $\beta_3^{\latt}$ means including the associated $\ln(N)$ running and fixing 
$c_2$ to its value \Eqre{CP}. We display this result in the second
column of Table~\ref{tab:PlaqCoeffserrors}. The shifts of the $p_n$
are well below the
statistical errors, even at high orders. It is worth mentioning
that the bulk of the changes is produced by fixing $c_1$ or $c_2$ to the
values \Eqre{CP}, while the different running is a subleading effect.
This explains why fixing $\beta_2=0$ had little impact on
the $p_n$-values: the $c_k$ ($k>1$) were kept as fit parameters.
Since the differences
between truncating at $\beta_1$-, $\beta_2$- or $\beta_3$-order
(see Table~\ref{tab:PlaqCoeffserrors}) can clearly be
neglected,
we take the differences between the results of the $\nu=9$ and $\nu=7$
fits displayed in Table~\ref{tab:PlaqCoeffsVol} as our systematic
uncertainties and add these in quadrature to the statistical
errors of our parameters from the primary $\nu=9$ fit. The final results are shown in the last column of
Table~\ref{tab:PlaqCoeffserrors}. All results from fits with acceptable
$\chired$-values that we performed, including those displayed
in the two tables, perfectly agree within errors with these
final results.

The above error analysis is quite similar to the one we did for
the expansion of the Polyakov line in Ref.~\cite{Bali:2013pla}.
In that case the systematic errors were dominant, and could
mainly be attributed to omitting higher $\beta$-function coefficients.
For the plaquette expansion the situation is quite different:
the systematic uncertainties are of the same size as the statistical
errors and are not dominated 
by the impact of omitting higher $\beta$-function coefficients.

The main parametric uncertainty in our case are
$1/N^6$-effects. Their significance should rapidly diminish
as the volume cut-off $\nu$ is increased. Therefore,
the systematic errors estimated above by varying $\nu$ should
also account for the truncation of the parametrization
at $\order(1/N^4)$.
We will now check
this assumption by adding $1/N^6$ corrections. 
As discussed in Sec.~\ref{sec:TBC},
we cannot include the most general $\order(1/N^6)$ expression
compatible with the OPE, which would require six additional
parameters for each order of the expansion.
Instead, we add the following simplified term:
\be
\label{N6term}
\delta \langle P \rangle_{\pert}(N)\sim
\frac{1}{N^6}
\sum_{n\geq 0}g_n\al^{n+1}((Na)^{-1})
\,.
\ee
This is expected to be the main contribution according to the
renormalon analysis of Sec.~\ref{sec:Renormalons} below.
This term introduces one new fit parameter per order of
the perturbative expansion and additional correlations between
different orders through the running of $\al((Na)^{-1})$. 
We perform this fit for different values of $\nu$ and display
the result obtained for $\nu=8$, which produced the smallest $\chired$-value,
in the first column of Table~\ref{tab:PlaqCoeffserrors}.
The differences between the central values of this and our primary fit
may be taken as estimates of the systematic errors associated
to the truncation of the parametrization at $\order(1/N^4)$.
We find these differences comparable in size to
those between the results of the $\nu=9$ and $\nu=7$ fits, without the
$1/N^6$ correction. 

\begin{figure}
\centerline{\includegraphics[width=0.9\textwidth]{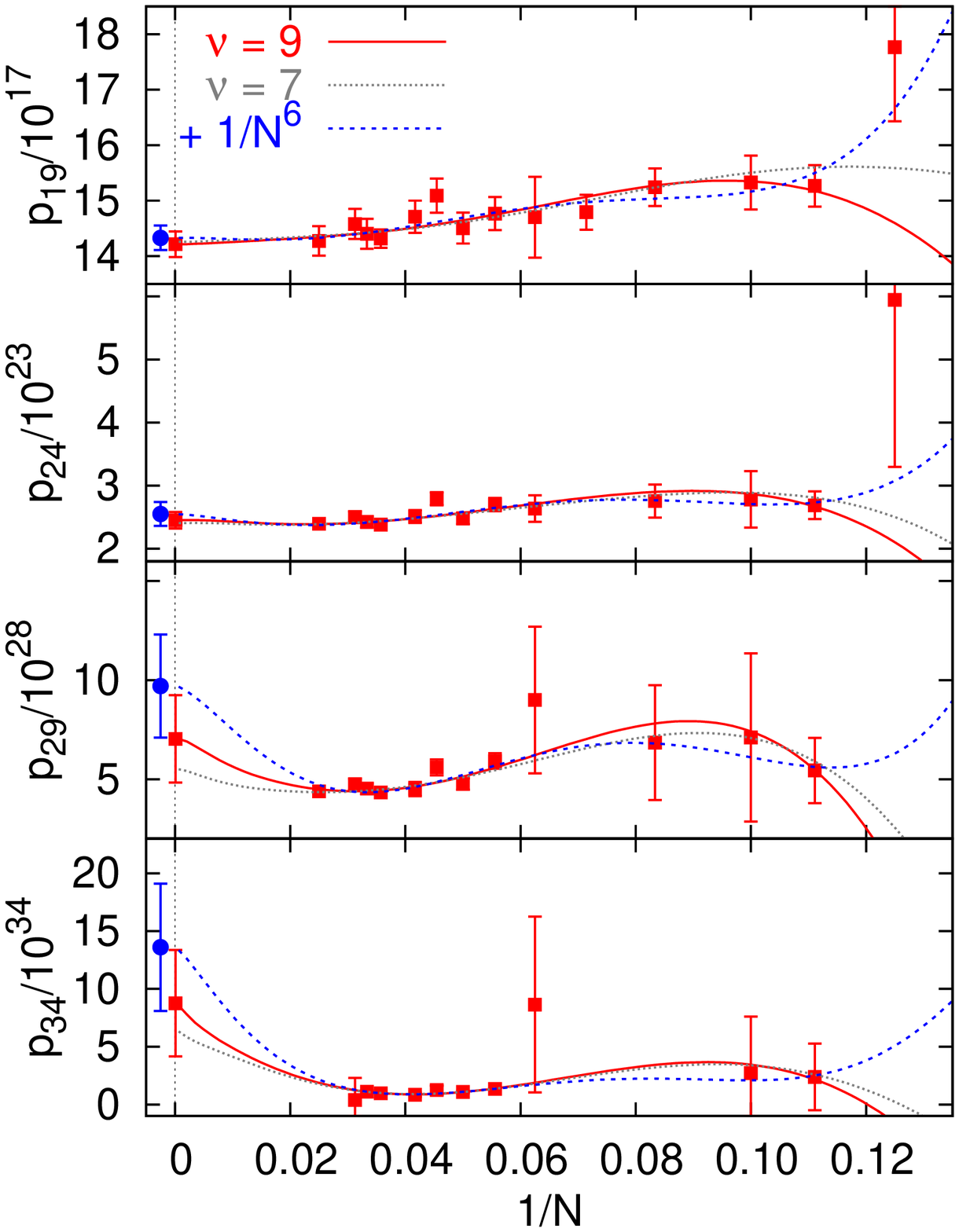}}
\caption{\it \label{fig:NLrunning} 
The TBC coefficients $p_{19,24,29,34}(N)$ as functions of $1/N$. The 
solid line represents the fit function
\Eqre{defaultfit} with $\nu=9$, the dotted line 
\Eqre{defaultfit} with $\nu=7$, and the dashed line \Eqre{defaultfit} plus the 
$1/N^6$-term \Eqre{N6term} with $\nu=8$. We also show (squares at $1/N=0$) our infinite volume extrapolations 
(last column of Table~\ref{tab:PlaqCoeffserrors}), as well as (circles) the infinite volume extrapolations including the 
$1/N^6$-term (first column of Table~\ref{tab:PlaqCoeffserrors}).}
\end{figure}

In Fig.~\ref{fig:NLrunning} we compare the NSPT finite volume data with
different fit functions for a few representative cases.\footnote{
We plot the data as a function of $1/N$ rather than of $1/N^4$,
to enhance the legibility. Otherwise all $N\geq 24$ points
would clutter in the very left of the figure.}
We plot our primary fit function \Eqre{defaultfit} with $\nu=9$
and with $\nu=7$, and the fit function including the $1/N^6$-effect \Eqre{N6term} with $\nu=8$.
We also show our final results for the infinite volume coefficients $p_n$ (last column of 
Table~\ref{tab:PlaqCoeffserrors}), as well as the results
from the fit including the $1/N^6$-effects (first column of 
Table~\ref{tab:PlaqCoeffserrors}).
From these figures the change of the curvature of the fit function
due to the running of $\al((Na)^{-1})$, that becomes
more pronounced as we increase the order $n$, is apparent.
The increase in curvature is expected from the asymptotic renormalon analysis,
see Sec.~\ref{sec:Renormalons} below.
We remark that the differences between our larger lattices,
i.e.\ $40\geq N\geq 24$,
and the values extrapolated to $N=\infty$
are much smaller here than they were in the case of the
Polyakov line~\cite{Bali:2013pla} where we went up to
$N(=N_S)=16$. 

\begin{table}[hb]
\ra{0.73}
\caption{\it Ratios $p_n/(np_{n-1})$
for different values of $\nu$
using the fit function \Eqre{defaultfit}, in analogy to
Table~\protect\ref{tab:PlaqCoeffsVol}.
The expectations from diagrammatic perturbation
theory are $p_1/p_0=1.278414323(14)$
and $p_2/(2p_1)=2.53952(9)$.}
\label{tab:RatioCoeffs}
\begin{ruledtabular}
\begin{tabular}{>{$}c<{$}|>{$}c<{$}>{$}c<{$}>{$}c<{$}>{$}c<{$}}
\nu & 13& 11&9 &7\\\hline
p_1/p_0                 & 1.27867(16) & 1.278506(60) & 1.278464(31) & 1.278434(28) \\
p_2/(2p_1)             & 2.53869(36) &  2.53929(17) &   2.53936(11) &  2.539406(95) \\
p_3/(3p_2)             & 2.21811(22) &  2.21794(15) &   2.21799(11) &    2.21803(10) \\
p_4/(4p_3)             & 1.91087(34) &  1.91088(21) &   1.91085(14) &    1.91088(12) \\
p_5/(5p_4)             & 1.66528(81) &  1.66549(62) &   1.66542(38) &    1.66540(20) \\
p_6/(6p_5)             &   1.4708(11) &  1.47078(98) &   1.47125(69) &    1.47110(29) \\
p_7/(7p_6)             &   1.3150(12) &    1.3150(10) &   1.31514(75) &    1.31547(35) \\
p_8/(8p_7)             & 1.18831(96) &  1.18824(86) &   1.18811(64) &    1.18865(35) \\
p_9/(9p_8)             & 1.08374(74) &  1.08367(69) &   1.08404(52) &    1.08406(42) \\
p_{10}/(10p_9)      & 0.99555(90) &   0.99559(82) &   0.99583(65) &    0.99624(59) \\
p_{11}/(11p_{10}) &   0.9204(15) &     0.9205(14) &     0.9209(11) &     0.92142(87) \\
p_{12}/(12p_{11}) &   0.8555(26) &     0.8556(21) &     0.8558(17) &       0.8554(14) \\
p_{13}/(13p_{12}) &   0.7996(34) &     0.7993(27) &     0.7998(22) &       0.8002(19) \\
p_{14}/(14p_{13}) &   0.7505(41) &     0.7499(33) &     0.7507(26) &       0.7510(23) \\
p_{15}/(15p_{14}) &   0.7069(49) &     0.7065(40) &     0.7071(31) &       0.7074(28) \\
p_{16}/(16p_{15}) &   0.6679(60) &     0.6680(50) &     0.6682(37) &       0.6684(33) \\
p_{17}/(17p_{16}) &   0.6330(73) &     0.6331(62) &     0.6339(44) &       0.6339(40) \\
p_{18}/(18p_{17}) &   0.6015(86) &     0.6020(74) &     0.6028(52) &       0.6029(47) \\
p_{19}/(19p_{18}) &     0.573(10) &     0.5758(88) &     0.5742(61) &       0.5743(54) \\
p_{20}/(20p_{19}) &     0.548(13) &       0.549(10) &     0.5495(70) &       0.5495(62) \\
p_{21}/(21p_{20}) &     0.526(15) &       0.528(11) &     0.5263(78) &       0.5260(70) \\
p_{22}/(22p_{21}) &     0.504(17) &       0.510(13) &     0.5063(87) &       0.5049(79) \\
p_{23}/(23p_{22}) &     0.485(19) &       0.495(14) &     0.4883(97) &       0.4847(90) \\
p_{24}/(24p_{23}) &     0.470(21) &       0.482(16) &       0.473(11) &       0.467(11) \\
p_{25}/(25p_{24}) &     0.463(25) &       0.475(20) &       0.461(14) &       0.452(13) \\
p_{26}/(26p_{25}) &     0.462(28) &       0.473(27) &       0.453(17) &       0.440(16) \\
p_{27}/(27p_{26}) &     0.462(33) &       0.478(38) &       0.450(22) &       0.430(20) \\
p_{28}/(28p_{27}) &     0.468(39) &       0.489(49) &       0.456(28) &       0.429(26) \\
p_{29}/(29p_{28}) &     0.485(49) &       0.505(59) &       0.471(34) &       0.438(32) \\
p_{30}/(30p_{29}) &     0.523(61) &       0.537(61) &       0.492(39) &       0.459(38) \\
p_{31}/(31p_{30}) &     0.512(63) &       0.527(60) &       0.509(37) &       0.489(37) \\
p_{32}/(32p_{31}) &     0.513(62) &       0.528(57) &       0.522(34) &       0.517(33) \\
p_{33}/(33p_{32}) &     0.503(65) &       0.519(57) &       0.527(31) &       0.535(31) \\
p_{34}/(34p_{33}) &     0.518(56) &       0.530(46) &       0.540(24) &       0.566(26) \\
\end{tabular}
\end{ruledtabular}
\end{table}

\begin{table}[hb]
\ra{0.73}
\caption{\it Ratios $p_n/(np_{n-1})$
for different fits. For details see the caption of
Table~\protect\ref{tab:PlaqCoeffserrors}.
In the last column we display the final values including their
statistical and systematic errors.}
\label{tab:RatioCoeffsErrors}
\begin{ruledtabular}
\begin{tabular}{>{$}c<{$}|>{$}c<{$}>{$}c<{$}>{$}c<{$}>{$}c<{$}}
                              &1/N^6& \beta_3\neq 0 &\beta_2=0    &{\rm Final\; result}\\\hline
p_1/p_0                 &  1.278554(54) & 1.278464(31) & 1.278464(31)  &1.278464(43)\\
p_2/(2p_1)             &  2.53916(17)  &   2.53936(11) &  2.53936(11)    &2.53936(12)\\
p_3/(3p_2)             &  2.21793(15)  &   2.21799(11) &    2.21799(11)  &2.21799(12)\\
p_4/(4p_3)             &  1.91081(22)  &   1.91086(14) &    1.91086(14)  &1.91085(15)\\
p_5/(5p_4)             &  1.66504(67)  &   1.66512(18) &    1.66542(39)  &1.66542(38)\\
p_6/(6p_5)             &  1.47077(99)  &   1.47174(48) &    1.47123(69)  &1.47125(70)\\
p_7/(7p_6)             &    1.3151(10)  &   1.31515(73) &    1.31515(76)  &1.31514(82)\\
p_8/(8p_7)             &  1.18827(81)  &   1.18802(63) &    1.18812(64)  &1.18811(83)\\
p_9/(9p_8)             &  1.08368(58)  &   1.08391(46) &    1.08404(52)  &1.08404(52)\\
p_{10}/(10p_9)      &  0.99541(74)  &   0.99568(62) &    0.99582(65) &0.99583(77)\\
p_{11}/(11p_{10}) &    0.9205(13)   &     0.9214(12) &     0.9209(11)   &0.9209(12)\\
p_{12}/(12p_{11}) &    0.8560(19)   &     0.8561(17) &       0.8558(17) &0.8558(17)\\
p_{13}/(13p_{12}) &    0.8000(25)   &     0.7997(22) &       0.7998(22) &0.7998(22)\\
p_{14}/(14p_{13}) &    0.7510(30)   &     0.7508(26) &       0.7507(26) &0.7507(26)\\
p_{15}/(15p_{14}) &    0.7077(36)   &     0.7069(31) &       0.7071(31) &0.7071(31)\\
p_{16}/(16p_{15}) &    0.6692(43)   &     0.6683(37) &       0.6682(37) &0.6682(37)\\
p_{17}/(17p_{16}) &    0.6346(51)   &     0.6343(44) &       0.6338(44) &0.6339(44)\\
p_{18}/(18p_{17}) &    0.6039(61)   &     0.6029(52) &       0.6027(52) &0.6028(52)\\
p_{19}/(19p_{18}) &    0.5761(72)   &     0.5743(61) &       0.5742(61) &0.5742(61)\\
p_{20}/(20p_{19}) &    0.5509(85)   &     0.5494(70) &       0.5495(70) &0.5495(70)\\
p_{21}/(21p_{20}) &    0.5277(98)   &     0.5257(79) &       0.5264(78) &0.5263(78)\\
p_{22}/(22p_{21}) &      0.508(11)   &     0.5056(87) &       0.5064(87) &0.5063(88)\\
p_{23}/(23p_{22}) &      0.492(14)   &     0.4876(96) &       0.4883(97) &0.488(10)\\
p_{24}/(24p_{23}) &      0.480(16)   &       0.471(11) &       0.473(11)   &0.473(12)\\
p_{25}/(25p_{24}) &      0.474(20)   &       0.459(13) &       0.461(14)   &0.461(16)\\
p_{26}/(26p_{25}) &      0.473(24)   &       0.449(15) &       0.453(17)   &0.453(22)\\
p_{27}/(27p_{26}) &      0.478(28)   &       0.445(19) &       0.451(22)   &0.450(30)\\
p_{28}/(28p_{27}) &      0.490(32)   &       0.449(24) &       0.457(29)   &0.456(39)\\
p_{29}/(29p_{28}) &      0.510(33)   &       0.463(29) &       0.472(34)   &0.471(47)\\
p_{30}/(30p_{29}) &      0.537(33)   &       0.484(34) &       0.494(39)   &0.492(51)\\
p_{31}/(31p_{30}) &      0.536(29)   &       0.502(33) &       0.511(37)   &0.509(42)\\
p_{32}/(32p_{31}) &      0.531(25)   &       0.515(29) &       0.524(34)   &0.522(34)\\
p_{33}/(33p_{32}) &      0.522(25)   &       0.521(28) &       0.529(31)   &0.527(32)\\
p_{34}/(34p_{33}) &      0.523(22)   &       0.537(23) &       0.542(24)   &0.540(35)\\
\end{tabular}
\end{ruledtabular}
\end{table}

We now determine the infinite volume $p_n/(np_{n-1})$-ratios.
These can be obtained from the same fits,
since we have also computed the correlation matrices.
The results for different values of $\nu$ using our default
fit function are displayed in Table~\ref{tab:RatioCoeffs}.
We find strong correlations between the errors of
consecutive expansion coefficients. Due to these correlations,
the infinite volume $p_n/(np_{n-1})$-ratios
are more precise than the coefficients themselves. We compute the central
values and the errors of the ratios
in the same way as we did for the coefficients. 
We show the results for the different variations
of the fit function we discussed above in Table~\ref{tab:RatioCoeffsErrors}.
Again in the last column we display our final numbers. For
the coefficients $p_n$ the statistical and systematic errors
were of similar magnitudes. In the case of the ratios
the total errors are dominated by statistics.
The systematics cancel to a large extent
and also the relative statistical uncertainties
are somewhat reduced, due to the above-mentioned
correlations between subsequent orders. 

Whereas we could determine the coefficients $p_n$ and their ratios
with reasonable accuracy, this is not the case for the $1/N^4$ correction
coefficients $f_n$ and $c_n$: these become compatible with zero within errors 
(albeit with central values significantly bigger than the $p_n$).
However, these parameters need to be included and their correlations 
are important to achieve acceptable fit qualities.

\section{Asymptotic behaviour of the expansion coefficients}
\label{sec:Renormalons}

In this section we confront the infinite volume coefficients $p_n$ obtained in Sec.~\ref{sec:fits}
with their large-$n$ dependence expected from the
renormalon picture. We start by presenting our theoretical expectations.
Then we compare these against the numerical data,
extract the normalization of the leading renormalon and
compare this with other determinations. We conclude estimating
the intrinsic ambiguity of truncated perturbative series.

\subsection{Renormalon analysis of the plaquette}
\label{sec:renpla}
The renormalon-associated large-$n$ dependence of
the coefficients $p_n$ means the perturbative expansion
of the plaquette is asymptotically divergent and its summation ambiguous.
This ambiguity is not arbitrary but 
such that it can be absorbed by
higher dimensional terms of the OPE, in our case by 
the gluon condensate $\langle O_{\G}\rangle$
(of dimension $d=4$) times its Wilson
coefficient $C_{\G}$ (see \Eqre{OPEMC}). 
This fixes the large-$n$ dependence of the $p_n$. 
Successive contributions to the sum $p_n\al^{n+1}$ should
decrease for increasing orders $n$ down to a minimum contribution for
$n_0 \sim 1/(a_d\beta_0)$, where $a_d=\beta_0/(2\pi d)$ (for a
more detailed discussion
see Sec.~\ref{sec:partial} below).
After this order the series starts to diverge.
Assuming the ambiguity of the sum to be of the order
of the minimum term we have $p_{n_0}\al^{n_0+1} \sim \exp[-1/(a_d\al)] \sim 
\lQ^da^d$, which can be absorbed redefining the gluon condensate. 

For notational convenience we introduce the following parametrization
of the integrated inverse $\beta$-function:
\be
\label{eq:betafun}
\Lambda=\mu\exp\left\{-\left[\frac{2\pi}{\beta_0\alpha(\mu)}
+b
\ln\left(\frac12 \frac{\beta_0\alpha(\mu)}{2\pi}\right)
+\sum_{j\geq 1}
s_j\,(-b)^j\!\left(\frac{\beta_0\alpha(\mu)}{2\pi}\right)^{\!j}\right]\right\}
\ee
with\footnote{Note that the $s_2$ we used in Ref.~\cite{Bali:2013pla}
equals $b(s_1^2/2-s_2)/(b-1)$ defined here.}
\begin{equation}
b=\frac{\beta_1}{2\beta_0^2}\,,\quad
s_1=\frac{\beta_1^2-\beta_0\beta_2}{4b\beta_0^4}\,,\quad
s_2=\frac{\beta_1^3-2\beta_0\beta_1\beta_2+\beta_0^2\beta_3}{16b^2\beta_0^6}\,.
\end{equation}
Note that the expansion coefficients $c_k$
defined in \Eqre{CP} are related to the above constants for the case of the Wilson action:
\begin{equation}
\label{eq:relatecs}
c_0=-b\frac{\beta_0}{2\pi}\,,\quad
c_1=s_1b\left(\frac{\beta_0}{2\pi}\right)^{\!2}\,,\quad
c_2=-2s_2b^2\left(\frac{\beta_0}{2\pi}\right)^{\!3}\,.
\end{equation}

The best way to quantify the asymptotic behaviour of the perturbative series is by
performing its Borel transform:
\be
B[P_{\pert}]\equiv\sum_{n=0}^{\infty}\frac{p_n}{n!}\left(\frac{4\pi}{\beta_0}u\right)^{\!n}\,.
\ee
The Borel transform of the expansion of the plaquette
will have a singularity, due to the dimension four gluon condensate,
at $u=d/2=2$:
\be
\label{eq:borel}
B[P_{\pert}]
=
N_P\, \frac{1}{(1-2u/d)^{1+db}}\left[1+b_1\left(1-\frac{2u}{d}\right)+
b_2\,\frac{db}{db-1}\left(1-\frac{2u}{d}\right)^2+\cdots\right]\,,
\ee
where (the second equalities apply to the Wilson action case)
\begin{align}
\label{eq:relatec0}
b_1&=ds_1+\frac{2\pi c_0}{\beta_0b}=ds_1-1\,,\\
\label{eq:relatec1}
b_2&=
\frac{4\pi^2c_1}{\beta_0^2b^2}
+ds_1\left(\frac{ds_1}{2}+\frac{2\pi c_0}{\beta_0b}\right)
-ds_2=ds_1\left(\frac{ds_1}{2}-1+\frac{1}{db}\right)-ds_2\,.
\end{align}

We skip the detailed derivation, which is quite standard (see, e.g.,
Ref.~\cite{Beneke:1998ui}),
and directly state the result of the Borel integral
for large orders $n$:
\begin{align}
\label{pn}
p_n &\stackrel{n\rightarrow\infty}{=} N_{P}\,
\left(
\frac{\beta_0}{2\pi d}\right)^{\!n}
\frac{\Gamma(n+1+db)}{\Gamma(1+db)}
\left\{
1+\frac{db}{n+db}\,b_1
\right.
\\\nn
&\qquad
\left.
+\frac{(db)^2}{(n+db)(n+db-1)}\,
b_2
+
\order\left(\frac{1}{n^3}\right)
\right\}
\,.
\end{align}
Note that the parameters $b_1$ and $b_2$
that describe the leading
pre-asymptotic corrections depend on
the expansion coefficients $c_0$ and $c_1$,
defined in \Eqre{CP},
of the Wilson coefficient
of the gluon condensate.

In the lattice scheme the numerical values
read\footnote{The error of
the $\order(1/n^2)$ coefficient is due to the uncertainty
of $\beta_3^{\latt}$, see \Eqre{eq:betas}.}
\begin{align}
\label{pnlatt}
p^{\latt}_n &\stackrel{n\rightarrow\infty}{=}
N^{\latt}_{P}\,\left(\frac{\beta_0}{2\pi d}\right)^{\!n}
\frac{\Gamma(n+1+db)}{\Gamma(1+db)}\\\nn
&\qquad\times \left\{
1+\frac{20.08931\ldots}{n+db}+\frac{505\pm 33}{\left(n+db\right) \left(n+db-1\right)}
+
\order\left(\frac{1}{n^3}\right)
\right\}
\,.
\end{align}
We observe that the pre-asymptotic corrections are quite large,
suggesting that high orders $n> 20$ are required 
to reach the asymptotic regime. Regarding this,
it is illustrative to show the corresponding
expansion in the $\MS$ scheme:
\begin{align}
p^{\MS}_n &\stackrel{n\rightarrow\infty}{=}
N^{\MS}_{P}\,\left(\frac{\beta_0}{2\pi d}\right)^{\!n}
\frac{\Gamma(n+1+db)}{\Gamma(1+db)}\\\nn
&\qquad\times \left\{
1-\frac{3.13653\ldots}{n+db}-\frac{1.1005\ldots}{\left(n+db\right) \left(n+db-1\right)}
+
\order\left(\frac{1}{n^3}\right)
\right\}
\,.
\end{align}
In this case the $1/n$ corrections are much smaller, suggesting
the asymptotic regime to be reached at much lower orders in the $\MS$ scheme
(as was seen in Ref.~\cite{Bali:2013pla}
for the expansion of the energy of a static source).

Note that $N_{P}$ dictates the strength of the
renormalon behaviour of any quantity where
the first non-perturbative effect is proportional to the
gluon condensate. Only the pre-asymptotic
effects will depend on the observable in question, due to different
Wilson coefficients. This motivates us to define
\be
N_{\G}=\frac{36}{\pi^2} N_{P}\,,
\ee
which is normalized in the same way as the gluon condensate.

$\langle P\rangle$ is a well-defined observable: it can be unambiguously computed 
in non-perturbative lattice simulations. Only after performing its OPE,
renormalon ambiguities show up.
They appear within individual terms of the OPE expansion but
have to cancel in the complete sum. 
\Eqre{pn} incorporates the leading renormalon behaviour 
of $P_{\pert}$, associated to the dimension four ($u=2$) matrix element. 
Dimension six ($u=3$) and higher order matrix elements in the OPE
will result in additional subleading renormalon
contributions to $P_{\pert}$. These, however, are
exponentially suppressed in $n$, relative to the leading renormalon,
and can be neglected.

More delicate, and of higher practical relevance, is the possible renormalon cancellation between dimension four and six matrix 
elements. This corresponds to a renormalon of dimension $6-4=2$ and implies that $C_{\G}(\al)$ may have a renormalon
itself to achieve this cancellation.
From the Borel plane point of view, we would then have 
\be
\label{eq:wilsonren}
B[C_{\G}] \sim \frac{1}{1-u}
\,.
\ee

Since the plaquette $a^{-4} P$ is a trivial
multiple of the Wilson gauge action Lagrange density, it can be related to
the trace anomaly:
\be
a^4 T^{\latt}_{\mu\mu} =
\frac{9\beta(\alpha)}{\pi \alpha^2} P 
\,.
\ee
This equality can be used to define the $\beta$-function
in the lattice scheme and this in turn allowed us to relate
the Wilson coefficient of the gluon condensate
$C_{\G}$ to the $\beta$-function
in \Eqre{CP}. Since each $c_k$-coefficient
contains a term proportional to $\beta^{\latt}_{k+1}$, 
the perturbative $\beta$-function will have a dimension
two infrared ambiguity, corresponding to a
renormalon at $u=1$. This can also be seen directly starting
from the expectation value
of the trace of the energy-momentum tensor \Eqre{eq:enmom}:
with the Wilson gauge action this equals $\langle T_{\mu\mu}^{\latt}\rangle$
up to $a^2\langle O_6\rangle$-type corrections. Defining
the $\beta$-function through the trace anomaly \Eqre{eq:enmom} then
results in the high-order behaviour 
of the coefficients $\beta_i$ to be
determined by a dimension two renormalon. Note that this does not imply that 
expansions of observables in terms of $\al(a^{-1})$
are affected by this singularity. However,
running $\al$ to a different scale will result in a $u=1$
divergent behaviour. This should not come as a surprise
since also in non-perturbative lattice simulations
masses etc.\ are subject to $\order(a^2\lQ^2)$
corrections under changes of the lattice scale $a$.
Note that the above arguments are specific for the
plaquette and the lattice scheme. We would not expect
the $\MS$ scheme $\beta$-function to receive renormalon
contributions.

We could be worried about the existence of ultraviolet
renormalons in the perturbative expansion of the plaquette, which
we have neglected in the above discussion. However, we do not
see any indication of alternating signs in the expansion of the
plaquette. Theoretically, this absence of ultraviolet renormalons
is expected since these can only appear when
integrating over momenta much bigger than the 
scale of $\al$. In our case this scale is $1/a$,
which is close to the maximum possible momentum $\sqrt{4}\pi/a$
that can be realized on a four dimensional lattice:
due to the hard cut-off perturbative expansions are
ultraviolet finite. 

Renormalons are not the only possible
sources of divergences. However, other singularities, e.g., due to
tunnelling instabilities are further
removed from the origin of the Borel plane.
For instance, instanton contributions
are suppressed by factors $\sim\exp(-2\pi/\alpha)$
for the case of TBC on symmetric lattices~\cite{'tHooft:1981sz,vanBaal:1982ag}
and, therefore, can only appear at $u\geq\beta_0/2\gg 2$.

Finally, for the ratios \Eqre{pn} implies
\begin{align}
\label{th:ratio}
\frac{p_n}{np_{n-1}}&=\frac{\beta_0}{2\pi d}
\left\{1
+\frac{db}{n}
+\frac{db(1-ds_1)}{n^2}
\right.\\\nn
&\qquad +\left.\frac{db\left[1-3ds_1+d^2b(s_1+2s_2)\right]}{n^3}+
\order\left(\frac{1}{n^4}\right)
\right\}\,.
\end{align}
The $1/n^2$- and $1/n^3$-correction terms
depend on the coefficients $c_0$ and $c_1$, which we eliminated
from the above equation
via \Eqre{eq:relatecs} (see also Eqs.~\eqref{eq:relatec0} and
\eqref{eq:relatec1}). We remark that
\Eqre{th:ratio} is a prediction, without any
free parameters, since $N_P$ cancels from the ratio.

\subsection{\boldmath Comparison to the numerical data}
\label{sec:Comparison}

The infinite volume extrapolation of the $p_n(N)$ made in Sec.~\ref{sec:fits} only 
used the OPE structure of the finite size effects. No assumption
was made about the divergent
behaviour of the perturbative series. We now compare
the extrapolated $p_n$-data with the renormalon-based expectations at large orders $n$.
We also determine the
normalization of the leading renormalon of the plaquette $N_P$
(and the associated one of the gluon condensate $N_{\G}=(36/\pi^2)N_P$) and
convert this into the $\MS$ scheme.

\begin{figure}[t]
\centerline{\includegraphics[width=1\textwidth,clip=]{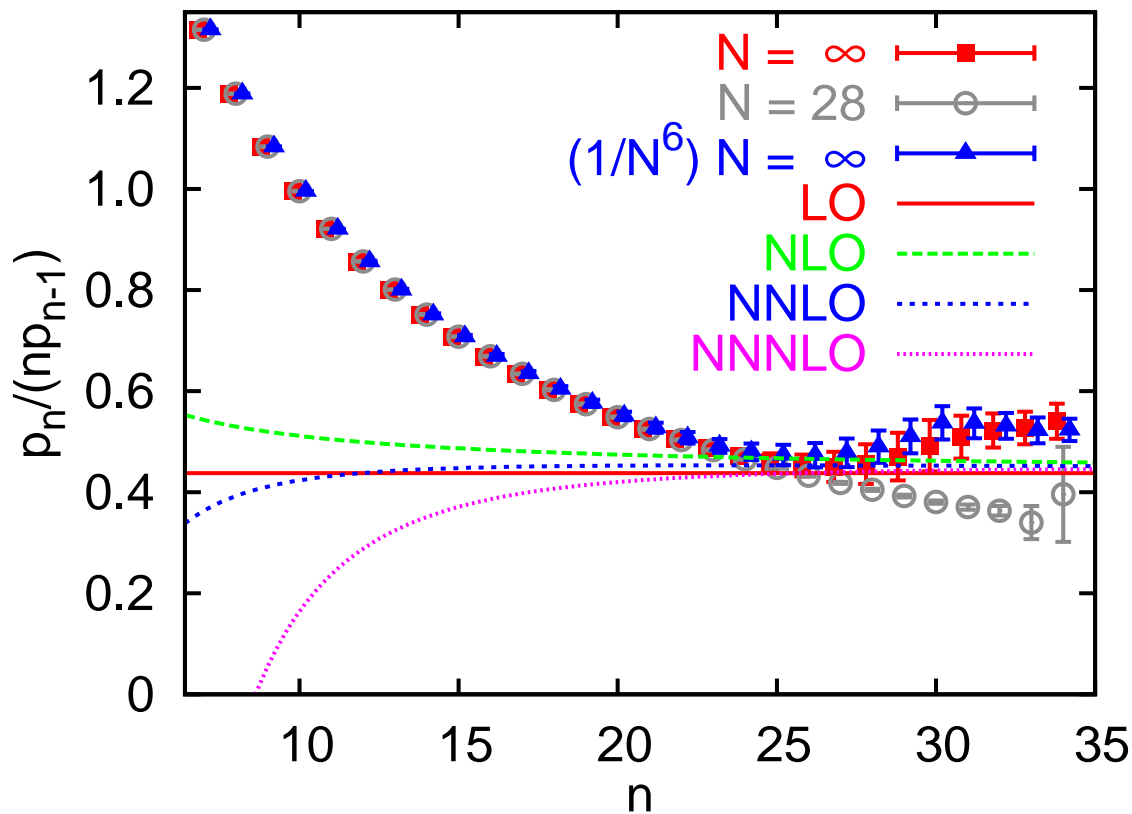}}
\caption{{\it The ratios
$p_n/(np_{n-1})$ compared with the prediction
\Eqre{th:ratio} for the LO,
next-to-leading order
(NLO), NNLO and NNNLO of the
$1/n$-expansion.
Only the ``$N=\infty$'' extrapolation includes the systematic uncertainties.
We also show finite volume data for $N=28$, and the result from the
alternative $N\rightarrow\infty$ extrapolation including $1/N^6$
corrections. The symbols have been shifted slightly horizontally.}
\label{n35}}
\end{figure}

In Fig.~\ref{n35} we compare our infinite volume
$p_n/(np_{n-1})$-ratios, summarized
in the last column of Table~\ref{tab:RatioCoeffsErrors},
to \Eqre{th:ratio}, truncating at different orders 
in the $1/n$-expansion.
As expected from the numerical values displayed in \Eqre{pnlatt},
we see quite substantial differences between the leading order (LO),
next-to-leading order (NLO), NNLO and
NNNLO curves. Therefore, in our Wilson lattice scheme,
we can only hope to detect the asymptotic behaviour for
orders $n\gtrsim 20$. Indeed, the data are in agreement
with the expectations for orders $n\geq 24$. For the highest
three orders ($n\geq 32$) the data are somewhat above the
expectation. However, these points are
highly correlated and at the very limit of what was achievable
for us, so we will not over-interpret this
behaviour.

In conclusion, the $p_n/(np_{n-1})$-ratios for $n\gtrsim 24$
clearly indicate the existence of a renormalon at $u=2$.
The coefficients
$p_n$ are certainly diverging and their asymptotic behaviour
is clearly inconsistent with other parametrizations, e.g.,
a singularity at $u=1$.
Unfortunately, we do not have enough precision to quantitatively
investigate subleading $1/n$-effects.

Next, we investigate the behaviour of the finite size effects. 
We expect the expansion coefficients of $\langle O_{\G}\rangle_{\soft}$,
i.e.\ the $f_n$ of \Eqre{eq:fnnn}, to
be governed by a dimension four ($u=2$) renormalon due to
its mixing with the Wilson coefficient of the unity operator,
i.e.\ $P_{\pert}$.
On a lattice with a fixed finite extent $N$
the divergence of the $f_n$ will, at very high orders,
result in an exponentiation of the associated logarithms, effectively
cancelling the $1/N^4$ suppression and the divergence of the $p_n$.
This will then, in the absence of non-perturbative terms,
result in a convergent expansion
of $\langle P \rangle_{\pert}(N)$. Therefore, finite size effects
are expected to become big for $n\gtrsim 24$. 
To illustrate this, we also display the finite volume $N=28$ data in Fig.~\ref{n35}. 
Indeed, for $n\gtrsim 24$, differences between the
$N=28$ data and the $N=\infty$ extrapolation become visible.
This is discussed in detail in Ref.~\cite{Bali:2013pla} for the
case of the expansion of the
static energy. In Eq.~(76) of this reference
$\beta_0$ needs to be replaced by $\beta_0/d$,
effectively quadrupling the order $n$ where this effect
becomes relevant, and $e^{\ln N_S}/N_S$ replaced by $e^{4\ln N}/N^4$ accordingly.
This behaviour also results in a more pronounced curvature of the fit function
at large $N$-values due to the running of $\al((Na)^{-1})$, as we increase the order $n$ (see Fig.~\ref{fig:NLrunning}). 
Nevertheless, for the plaquette, these running effects get obscured by the $u=1$ renormalon 
of the Wilson coefficient
$C_{\G}$, since the $c_k$ saturate towards the asymptotic behaviour at lower orders
than the $f_n$ and then diverge more rapidly
($c_k\sim k\beta_0/(4\pi)c_{k-1}$ rather than
$f_{n}\sim n\beta_0/(8\pi)f_{n-1}$). However, in this
asymptotic regime the $1/N^6$ coefficients $g_n$ are also
expected to diverge, the associated logarithms to exponentiate
and to cancel against the $c_k/N^4$- and $p_n$-contributions.

Our fits are consistent with the above picture. 
We expect that our primary fit, which does not
incorporate $\order(1/N^6)$ terms, only provides an effective parametrization
of $1/N^4$ and $1/N^6$ renormalon-associated effects. 
We first observe that setting the Wilson coefficient $C_{\G}$
to one, i.e.\ $c_k=0$, we cannot simultaneously account for the
$u=2$ renormalon of the $f_n$ parameters and for the
effects of the $u=1$ renormalon on the $c_k$ parameters.
Within our primary fit we observe
the central values of the parameters $f_n$ and $c_k$ to grow much faster
towards high orders than the $p_n$-coefficients. This is consistent
with the existence of a $u=1$ renormalon since, 
in the absence of $1/N^6$-terms, cancellations have to
take place between combinations of $f_n$- and $c_k$-terms. In any case, 
we remark that the individual coefficients all carry large relative errors
of $\order(1)$. Therefore, these statements
are qualitative in nature rather than quantitative.
A reliable determination
of the $c_k$- and $f_n$-coefficients (and of their expected divergences)
requires a full $\order(1/N^6)$ analysis,
with six additional fit parameters per order of the expansion,
which is beyond our reach. Instead, we partially included the leading
$\order(1/N^6)$ logarithms into our fits according to \Eqre{N6term}. 
As a result, the growth of the $c_k$-coefficients becomes more consistent
with a $u=1$ renormalon. Also the $g_n$-values are observed to
grow much faster towards high orders than the $p_n$-coefficients.
The coefficients $f_n$ are comparatively smaller in size than the $c_n$ and 
$g_n$ but larger than the
corresponding $p_n$. Also in this case, all the finite size
coefficients carry large relative errors
of $\order(1)$, making this discussion, at most, qualitative.

Fortunately, for the coefficients $p_n$ the $1/N^6$-effects
are only subleading and, as can be read off from
Table~\ref{tab:PlaqCoeffserrors}, their values change very little
when adding some of these higher order effects.
The errors of our infinite volume coefficients
$p_n$ in the last column of Table~\ref{tab:PlaqCoeffserrors}
already incorporate these systematics. We illustrate this by 
including the extrapolation to infinite $N$, incorporating a $1/N^6$-term
(first column of Table~\ref{tab:RatioCoeffsErrors}),
into Fig.~\ref{n35}.
The errors displayed in this case are only statistical.

It is worth mentioning that
in the case of the static energy studied in
Refs.~\cite{Bauer:2011ws,Bali:2013pla,Bali:2013qla} the Wilson
coefficient of the leading (in this case $d=1$) finite volume correction
was exactly one. Consequently, there were no ambiguities that had to be 
absorbed by even
higher dimensional operators.
Therefore, the above complication was not encountered
and we were not only able to reliably determine the infinite
volume expansion coefficients but also the coefficients
of the $1/N$ finite volume correction term.

\subsection{\boldmath Determination of $N_P$}
\label{sec:NP}

\begin{figure}
\centerline{\includegraphics[width=0.98\textwidth,clip]{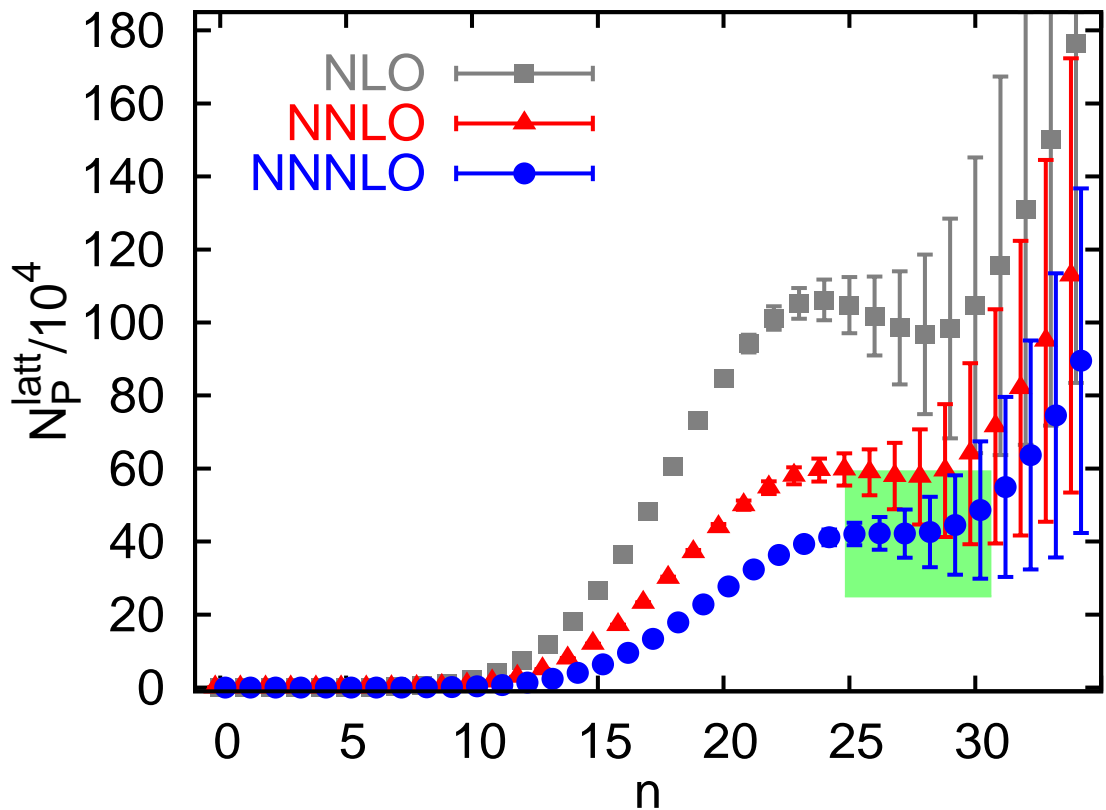}}
\caption{{\it $N_P$, determined from the
coefficients $p_n$ via \Eqre{pn}
truncated at NLO, NNLO and NNNLO. The green box
marks our final result quoted in 
\Eqre{NPfinal}. The data are slightly adjusted horizontally.}
\label{fig:NP}}
\end{figure}

To obtain the normalization $N_P$ we divide the
coefficients displayed in
Table~\ref{tab:PlaqCoeffserrors} by \Eqre{pn}
truncated at different orders in $1/(n+db)$, labelled
as (for consistency with \Eqre{th:ratio} and Fig.~\ref{n35}) NLO, NNLO and NNNLO, respectively.
For large $n$-values these ratios should tend to constants, allowing us
to extract $N_{P}$. This is depicted in
Fig.~\ref{fig:NP}. We observe the three data sets
are compatible with constant values for $n\gtrsim 24$.\footnote{In
the case of the static energy we obtained an extremely
clear plateau within small errors~\cite{Bauer:2011ws,Bali:2013pla,Bali:2013qla}.
Unfortunately, in the present case
the errors grow quite rapidly for $n\gtrsim 30$.}
In Fig.~\ref{fig:NP} we also observe that truncating 
\Eqre{pn} at different orders in $1/(n+db)$ produces large
corrections. Fortunately enough, however, they follow a convergent pattern, 
with smaller differences between the
NNLO and NNNLO curves than between the NLO and NNLO curves.
We also note that in the range $25\leq n\leq 30$, where we regard
the prediction as most reliable, the inclusion of higher order $1/n$ 
effects results in a flatter dependence on $n$.

We take the value of the NNNLO evaluation 
for $n=26$, where it exhibits a very mild maximum, as our central
value.
For $n<25$ we may not have reached the asymptotic behaviour
whereas for $n>30$ the results become less meaningful, due to the
exploding errors. The uncertainty of the determination
of $N_P$ is dominated by the pre-asymptotic effects, which
are large in the lattice scheme. We
use the difference between the NNNLO and NNLO determinations at $n=26$
as an estimate for even higher order effects
and add this in quadrature to the (comparatively small) error of the NNNLO
prediction:\footnote{Any other value within the range $25\leq n\leq 30$ 
agrees with \Eqre{NPfinal} within the error. This is a
reflection of strong correlations between the data.}
\begin{align}
 N_P^{\latt}=42(17)\times 10^4 \,,&\qquad N^{\latt}_{\G}=1.54(63)\times10^6\,,\nn\\
 N_P^{\MS}=0.61(25)\,,&\qquad N^{\MS}_{\G}=2.24(92)\,.
\label{NPfinal}
\end{align}
For the last two equalities we have used the exact identity
\be
N^{\MS}_{P}=N^{\latt}_{P}\Lambda^4_{\latt}/\Lambda^4_{\MS}\,,
\ee 
where~\cite{Hasenfratz:1980kn,Luscher:1995np}
$\Lambda_{\MS}\approx
28.809338139488\,\Lambda_{\latt}$.
Note that the normalization of the plaquette renormalon
in the $\MS$~scheme is of $\order(1)$, 
as it is the case for the renormalon of the heavy quark pole
mass [cf. Eq.~(105) of Ref.~\cite{Bali:2013pla}, or Eq.~(11) of Ref.~\cite{Bali:2013qla}]. 

We have also explored alternative methods to determine $N_P$. One
is using the relation
\be
\label{NPBorel}
N_P=\left.B[P_{\pert}](u)(1-2u/d)^{1+bd}\right|_{u=2}
\ee
to compute $N_P$ as a perturbative expansion in $u$ \cite{Lee:1996yk}.
However, this did not work, which may not be surprising since
the singularity is located at $u=2$, very far away from the origin.
One may also consider a conformal mapping to move the singularity
closer to the origin. Again, we do not obtain the expected plateau
behaviour for the orders of the expansion that we have at our
disposal. This is consistent
with the analysis made in Ref.~\cite{Bali:2013pla}, where this
method became compatible with the asymptotic expectation only
at much higher orders (compare Fig.~12 with Fig.~14 of this reference)
than the method we outlined and employed above. In Ref.~\cite{Bali:2013pla}
we were able to go to orders $(n+1)/d\leq 20$ rather than $(n+1)/d\leq 35/4$
and ultimately found agreement between the
two determinations.

We now compare \Eqre{NPfinal} with previous estimates
available in the literature. The large-$\beta_0$ 
result can be found, for
instance, in Refs.~\cite{Broadhurst:1992si,Beneke:1998ui}:
\be
\label{NPbeta0}
N^{\MS}_{P,\mathrm{large-\beta_0}}=\frac{e^{10/3}}{24\pi}\approx 0.37178
\,.
\ee
This is 40\% smaller than our central value but within
errors still consistent with our result $N_P^{\MS}=0.61(25)$.\footnote{Note though that a different definition of the
Borel transform  $\sim \tilde{N}_P/(a-2ua/d)^{1+db}\cdots$
in Eq.~(\ref{eq:borel}) would introduce arbitrary
factors $a^{db}$, relative to this large-$\beta_0$ result. We thank Matthias Jamin for discussions on this point.}
There also exist
estimates from the perturbative expansion of the Adler function. 
In Ref.~\cite{Beneke:2008ad} the first four orders
were used to fit the expected leading renormalon singularities
in the Borel plane (see also the discussion in Ref.~\cite{Beneke:2012vb}).
The result was $N^{\MS}_{P}\approx 0.02$ for $n_f=3$. For the case of
$n_f=0$, which corresponds to our setting, this model
yields~\cite{Matthias} $N^{\MS}_{P}\approx 0.04$
(note the strong dependence on $n_f$). In Ref.~\cite{Lee:2011te}
the value 0.01 was obtained using the conformally mapped version
of~\Eqre{NPBorel} for the Adler function.
We remark that using the method of Ref.~\cite{Lee:2011te}
we were not able to obtain
the renormalon normalization with our $\order(\al^{35})$ perturbative
expansion.
While these numbers differ quite substantially from each other,
all of them are significantly smaller than our determination.
We believe that the main difficulty with these analyses is
that the perturbative expansion of the Adler function is not known
to sufficiently high orders to probe the $u=2$ renormalon.
Also in our case, see Fig.~\ref{fig:NP}, lower orders would have
given smaller numbers. While it should not be necessary to
go up to $n>20$ to detect the renormalon in the $\MS$ scheme,
also in this case orders four times higher than for the
heavy quark pole mass renormalon at $u=1/2$ probably are necessary.

\subsection{Partial sum and minimal term}
\label{sec:partial}
In the regime where the coefficients $p_n$ are
dominated by the renormalon behaviour, we can 
determine the order
$n_0$ that corresponds to the minimal term
of the perturbative series from
the analytical expectation \Eqre{pn}.
Minimizing $p_n\al^{n+1}$ results in 
\be
\label{minimizing}
(n_0+db)\frac{\beta_0\al}{2\pi d}=\exp\left\{-\frac{1}{2(n_0+db)}+\order\left[\frac{1}{(n_0+db)^2}\right]\right\}\,.
\ee
This then gives the minimal term
\begin{align}
\label{minterm}
p_{n_0}\alpha^{n_0+1}
&=\frac{2\pi d^{1/2+db}}{2^{db}\Gamma{(1+db)}}\sqrt{\frac{\alpha}{\beta_0}}N_P
\exp\!\left(-
\frac{2\pi d}{\beta_0\alpha}\right)\left(\frac{\beta_0\alpha}{4\pi}\right)^{-db}
\left[1+\order(\al)\right]\\\nn
&\approx
\frac{2\pi d^{1/2+db}}{2^{db}\Gamma{(1+db)}}\sqrt{\frac{\alpha}{\beta_0}}(\Lambda a)^4
\,.
\end{align}

While the perturbative series is divergent,
truncating it at the order $\nmax\simeq n_0(\alpha)$,\footnote{In practice
one would round $\nmax=\mathrm{int}(n_0+1/2)$.}
\be
S_P(\al)=\sum_{n=0}^{n_{\max}}p_n\al^{n+1}
\,,
\ee
results in a finite sum
(this is equivalent to a particular scheme to subtract the renormalon).

By taking $\nmax\simeq n_0$ we minimize the dependence
of the series on the order at which it is truncated.
We assign the uncertainty of the sum due to the truncation to be
\be
\label{minterm2}
\delta S_P = \sqrt{n_0}\,p_{n_0}\alpha^{n_0+1}\approx
\frac{(2\pi)^{3/2}d^{1+db}}{2^{db}\beta_0\,\Gamma{(1+db)}}N_P(\Lambda a)^4
\approx 12.06\, N_P(\Lambda a)^4
\,.
\ee
This object is scheme- and
scale-independent (to the $1/n$-precision that we employed in the
above derivation) because, even though the normalization $N_P$ 
depends on the scheme, the product $N_P\Lambda^4$
is scheme-independent. A higher order calculation
should yield an expression that is proportional to 
the product of \Eqre{minterm2} and the Wilson coefficient $C_{\G}$,
since the ambiguity of the truncated sum must cancel
against a similar ambiguity of
the contribution from the gluon condensate.

\begin{figure}
\centerline{\includegraphics[width=.99\textwidth,clip]{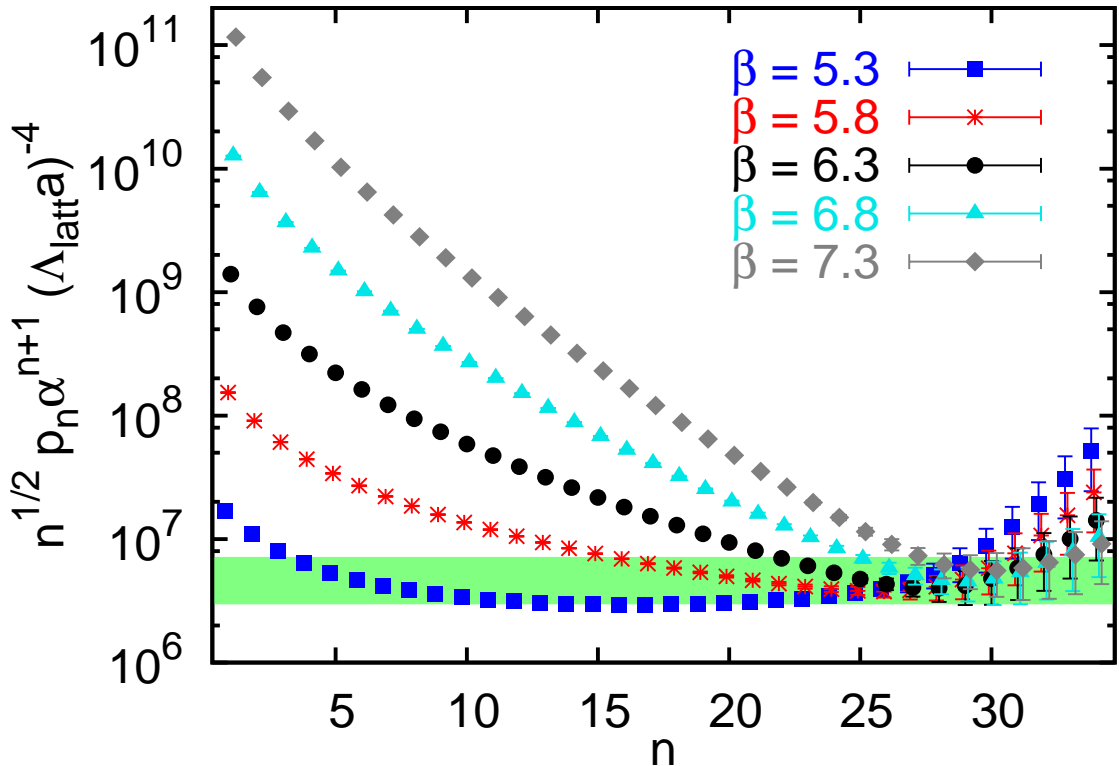}}
\caption{{\it The combination $\sqrt{n}\,p_n\alpha^{n+1}/(\Lambda_{\latt}a)^4$,
see \Eqre{eqplot}, as a function of
$n$ for $\beta=5.3, 5.8, 6.3, 6.8$ and 7.3.
The error band corresponds to the theoretical expectation
$12.06\,N_P=5.1(2.1)\times 10^6$ of \Eqre{minterm2}, where we have used 
the value of \Eqre{NPfinal} for $N_P$. The data sets have been
adjusted horizontally for better legibility. Note that the left-most points
correspond to $n=1$.}
\label{fig:minterm}}
\end{figure}

In Fig.~\ref{fig:minterm} we plot the combination
\be
\label{eqplot}
\frac{\sqrt{n}\,p_n\alpha^{n+1}}{(\Lambda a)^4}
\approx
\sqrt{n}\,p_n\alpha^{n+1}
\displaystyle{
e^{
4\left[\frac{2\pi}{\beta_0\alpha}
+b
\ln\left(\frac12 \frac{\beta_0\alpha}{2\pi}\right)
-s_1b\frac{\beta_0\alpha}{2\pi}
+s_2b^2\left(\frac{\beta_0\alpha}{2\pi}\right)^{\!2}
\right]}
}
\ee
as a function of
$n$ where
we substitute $1/(\Lambda a)^4$ by the 
integrated four-loop $\beta$-function
of \Eqre{eq:betafun}.
For $n\simeq n_0$ 
\be
\label{eqplot2}
\left.\frac{\sqrt{n}\,p_n\alpha^{n+1}}{(\Lambda a)^4}\right|_{n=n_0}
=\frac{\delta S_P}{(\Lambda a)^4}
\simeq
12.06\,N_P
\,,
\ee
so it should approach the value
$12.06\, N_P=5.1(2.1)\times 10^6$ [\Eqre{minterm2}
with the $N_P$-value of \Eqre{NPfinal}], drawn as an error band.
The comparison is made for
$\beta=3/(2\pi\al)=5.3, 5.8, 6.3, 6.8$ and $7.3$.
The three values $\beta=5.8, 6.3$ and 6.8 are
typical for present-day non-perturbative lattice
simulations, with inverse lattice spacings 
$1.4 \,\mathrm{GeV}\lesssim
a^{-1}\lesssim 
6.4\,\mathrm{GeV}$~\cite{Necco:2001xg}, while
$\beta=5.3$ is in the strong-coupling regime.

The corresponding $n_0$-predictions \Eqre{minimizing} are,
in ascending order of the $\beta$-values
$n_0\simeq 24, 26, 28, 30$ and 33.
In the figure we have
multiplied the minimal term
by $\sqrt{n}$ which then corresponds to the
uncertainty of the truncated series.
Note that the variation
of $\sqrt{n}$ for $24\leq n\leq 33$ can be neglected
on the logarithmic scale of the figure.
As expected, the contributions to the sum decrease
monotonously down to the minimum at orders
that, within errors and for $\beta\geq 5.8$, are
consistent with the above expectations on $n_0$.
Thereafter, the contributions start 
to diverge.\footnote{The exponential divergence was more
clearly observed for the static energy (see Fig.~15 of
Ref.~\cite{Bali:2013pla}), where the divergence is expected
to be stronger ($u=1/2$)
and where we were also able to compute a higher number
of orders with $n>n_0$.}
The ambiguity computed from the data agrees perfectly
with the prediction. This is quite remarkable, as the sizes
of the different terms of the perturbative series cover
several orders of magnitude.

The effect of truncating the integrated
$\beta$-function \Eqre{eq:betafun} at different orders 
in \Eqre{eqplot} is sizeable because $|\beta_2^{\latt}|$ and 
$|\beta_3^{\latt}|$ are numerically large
and $d=4$. The $1/(n+db)$- and $1/(n+db)^2$-terms of \Eqre{pn}
(for numerical values see \Eqre{pnlatt})
have the same origin. Including the $s_1\al$- or $s_2\al^2$-terms
into \Eqre{eqplot}
has a similar effect as the inclusion of the $1/n$- or $1/n^2$-terms
had on the determination of the normalization
$N_P$, see Fig.~\ref{fig:NP}.
Therefore, best agreement is achieved truncating
\Eqre{eqplot} at the order in $\al$ associated to the
respective $1/n$ truncation of the $N_P$-determination.
The Wilson coefficient $C_{\G}$ that we have
ignored so far would reduce the data points
by only a few per cent within the range of couplings covered
by the figure and can safely be neglected.

It is interesting to see that the order at which
the series starts exploding can be delayed by
decreasing the coupling, i.e.\ going to larger $\beta$-values,
however, the ambiguity of the expansion remains the same since
its origin lies in the inherent ambiguity of the
definition of the non-per\-tur\-ba\-tive gluon condensate.
We estimate this ambiguity
using the result \Eqre{NPfinal} for $N_{\G}^{\MS}$, the
prefactor of \Eqre{minterm2}
and the value~\cite{Capitani:1998mq,Necco:2001xg}
$\Lambda_{\MS}(n_f=0)= 0.602(48) r_0^{-1}\approx 238~\mathrm{MeV}$:
\be
\label{eq:ambi2}
\delta \langle O_{\G}\rangle_{\NP}
\simeq
\left.\frac{(2\pi)^{3/2}d^{1+db}}{2^{db}\beta_0\,\Gamma{(1+db)}}\,N_{\G}^{\MS}\right|_{n_f=0}\Lambda_{\MS}^4=27(11)\,\Lambda_{\MS}^4
 \sim 0.087 \,\mathrm{GeV}^4\,.
\ee
$n_f=0$ relates to the $n_f$-dependence of $\beta_0$ and $b$.
The above value is bigger than standard 
estimates of the non-perturbative gluon
condensate~\cite{Vainshtein:1978wd}
$\sim 0.012\,\mathrm{GeV}^4$, and
indicates that determinations of this quantity
may significantly depend on the way the perturbative 
series is truncated or approximated. Note that the large-$\beta_0$ limit of 
\Eqre{eq:ambi2} (using \Eqre{NPbeta0}) yields a 
considerably smaller number, which, however, is still bigger than standard 
estimates:
\be
\label{eq:ambibeta0}
\delta \langle O_{\G}\rangle_{\NP,\mathrm{large-\beta_0}}
\simeq
\left.\frac{(2\pi)^{3/2}}{\beta_0}
\,
\frac{6e^{10/3}}{\pi^3}\right|_{n_f=0}\Lambda_{\MS}^4
\simeq
7.77\,\Lambda_{\MS}^4
 \sim 0.025 \,\mathrm{GeV}^4\,.
\ee

Eqs.~\eqref{eq:ambi2} and \eqref{eq:ambibeta0} have been computed for $n_f=0$.
The prefactors multiplying $N_{\G}^{\MS}$ 
only show a mild $n_f$-dependence in both cases.
While the large-$\beta_0$ limit of $N_{\G}^{\MS}$ is $n_f$ independent, 
beyond this approximation the $n_f$-dependence of $N_{\G}^{\MS}$
is unknown.

\section{Summary and Conclusions}

The expectation value of the (infinite volume) plaquette
can be expanded as follows
\be
\langle P\rangle=P_{\pert}(\al)\langle\mathds{1}\rangle
+a^{4}\frac{\pi^2}{36}C_{\G}(\al)\langle O_{\G}\rangle_{\NP}+\order(a^6)\,,
\ee
where $\langle O_{\G}\rangle_{\NP}$ is the renormalization group
invariant 
definition of the non-perturbative gluon condensate
and $C_{\G}(\alpha)=1+\order(\al)$ is its Wilson coefficient.
In our numerical stochastic
perturbation theory simulation, we calculated the coefficients $p_n(N)$ of the perturbative expansion
\be
\langle P \rangle_{\pert}(N)=\sum_{n\geq 0}p_n(N)\al^{n+1}\,
\ee
in
lattice regularization with the Wilson gauge action
up to $\order(\al^{35})$  on
lattices of up to $40^4$ points, using twisted boundary
conditions (TBC) in three directions. The choice of
TBC turned out to be superior to periodic boundary
conditions, not only in terms of statistical
errors and reduced finite volume effects, but also because only these
boundary conditions allow for a systematic analysis of
finite volume effects in the framework of the
operator product expansion (OPE). This enabled us to accurately obtain the 
infinite volume extrapolation of the $p_n$-coefficients: 
\be
P_{\pert}=\lim_{N\rightarrow\infty}\langle P \rangle_{\pert}(N) \qquad {\rm and} \qquad 
p_n=
\lim_{N\rightarrow\infty}p_n(N)\,,
\ee
as well as of their ratios $p_n/(np_{n-1})$.
The results are summarized in the last columns
of Tables~\ref{tab:PlaqCoeffserrors} and 
\ref{tab:RatioCoeffsErrors}. We have analysed their high-order behaviour 
and found the $p_n$-coefficients to diverge from orders $n\gtrsim 24$ onwards
in a way consistent with a renormalon at $u=2$ in the Borel
plane, as expected from the dimensionality $d=4$ of the gluon
condensate. This is illustrated in Fig.~\ref{n35}. We stress that
we were only able to obtain this result after having achieved both
good theoretical control of finite volume effects
and computing the perturbative expansion to orders as high as $\al^{35}$.

Furthermore, we have determined the normalization $N_P$ of the corresponding renormalon
(see Eqs.~\eqref{eq:borel} and \eqref{pn} for its definition): 
\begin{align}
 N_P^{\latt}=42(17)\times 10^4 
 \,.
 \end{align}
This can be converted from the lattice into the
$\MS$ scheme at arbitrary precision since the combination
$N_P\Lambda^4$ is scheme-independent.
We obtained $N_P^{\MS}=0.61(25)$ in the $\MS$ scheme, 
which differs by 2.5 standard deviations from zero.
Still, a 40\% error on $N_P\Lambda^4$ translates
into a 10\% error on the $d=1$ combination
$N_P^{1/4}\Lambda$. Alternatively, we can normalize the series accompanying
$\langle\mathds{1}\rangle$
consistently with respect to $\langle O_{\G}\rangle$,
to obtain the normalization of the renormalon
associated to the gluon condensate:
\be
\label{NPcon}
N^{\MS}_{\G}=\frac{36}{\pi^2}N_P^{\MS}=2.24 \pm 0.92\,.
\ee
This is independent of any pre-asymptotic effects
and therefore of the observable in question.
From this value we can also estimate the intrinsic truncation ambiguity of
corresponding perturbative expansions,
see Eqs.~\eqref{minterm2} and \eqref{eq:ambi2},
\be
\delta \langle O_{\G}\rangle_{\NP}
\simeq (27\pm 11)\,\Lambda_{\MS}^4\,.
\ee
This is larger than standard
estimates of the non-perturbative gluon condensate~\cite{Vainshtein:1978wd}
$\sim 0.012$. Therefore, determinations of this quantity
may significantly depend on the way the perturbative
series is truncated or approximated. 
The above value is by a factor of 3.5 bigger than the
large-$\beta_0$ result and 
by about one order of magnitude larger than many previous estimates of the
ambiguity of the gluon condensate, see, for instance,
Eq.~(5.12) of Ref.~\cite{Beneke:1998ui}. This is mainly
due to the large prefactor relating $N_P$ to
$\delta S_P$ in \Eqre{minterm2}, and $N_{\G}$ to
$\delta \langle O_{\G}\rangle_{\NP}$ in \Eqre{eq:ambi2}. 
Finally, we remark that we obtain a similar uncertainty
just by computing $\sqrt{n_0}p_{n_0}\al^{n_0+1}$ directly
from the data, see Fig.~\ref{fig:minterm}, thereby verifying
this large prefactor.

The magnitude of pre-asymptotic
$1/n$- and $1/n^2$-corrections was the
main limiting factor for the precision of
\Eqre{NPcon}.
In our case, we suffered from large coefficients
$|\beta_2^{\latt}|$ and $|\beta_3^{\latt}|$. This is not the case 
in the $\MS$ scheme. Actually, there are strong indications
(see, e.g., Ref.~\cite{Pineda:2001zq})
that renormalon dominance for the pole mass in the $\MS$ scheme sets in
already at orders as low as $n\lesssim 2$. Therefore, in this scheme
perturbative expansions of observables with non-perturbative
contributions from $\langle O_{\G}\rangle_{\NP}$
may show the expected asymptotic behaviour already for orders
$n\lesssim 8\ll 24$. However, a direct translation of the
perturbative coefficients
from the lattice to the $\MS$ scheme is not possible, 
since the necessary conversion is not known to such high orders. 
In Ref.~\cite{Bali:2013pla} we experimented with
resumming the expansion by re-defining the coupling,
without changing the action or observable, so that it resembled
a $\MS$-like scheme.
This resulted
in an earlier on-set of the asymptotic behaviour, however,
at the price of much larger statistical errors so that
the determination of the normalization could not be improved upon.
Alternatively, it is conceivable that other lattice discretizations
with smaller $\Lambda_{\MS}/\Lambda_{\latt}$-ratios will have smaller
high-order $\beta$-function coefficients,
resulting in renormalon dominance at smaller orders $n$. 
In particular, the $\order(a^2)$ Symanzik-improved
action~\cite{Weisz:1982zw,Luscher:1984xn} would be worthwhile to study.
Unfortunately, in this case fewer analytic and semi-analytic
low-order results are available. Finally, we would also like to 
stress that pre-asymptotic effects
do not only depend on the $\beta$-function coefficients
but also on $C_{\G}$.
Therefore, the on-set of renormalon dominance depends
both on the renormalization scheme and on the
observable in question. 

Our analysis may immediately impact on phenomenological
analyses in cases where the perturbative series is sensitive to the
gluon condensate renormalon. Even though one should bear in mind
that we have only studied the pure gauge $n_f=0$ theory,  
it is worth mentioning that for the pole mass renormalon ($u=1/2$) 
the $n_f$ dependence has been found to be mild.
In that case an analysis analogous to the one
performed in the present paper yielded 
a precision of 6\% for the associated normalization
$N_m\Lambda$~\cite{Bali:2013qla} for the $n_f=0$ theory.
The resulting value was only 8\%  
off of the $n_f=3$ result obtained in Ref.~\cite{Pineda:2001zq} from the pole mass perturbative expansion (up to orders $n=3$)
in the $\MS$ scheme. It is also reassuring that the $n_f$-dependence of the large-$\beta_0$ result is under control (with a difference 
 of $\sim 20$\% between the $n_f=3$ and $n_f=0$ results of
\Eqre{eq:ambibeta0}). 
In any case, it would certainly be worthwhile
to repeat our investigation using a different gauge
action and incorporating fermions. Such future
studies will not change, however, the qualitative picture or the
main conclusions presented here.

\begin{acknowledgments}
We thank V.~Braun, M.~Golterman and M.~Jamin for discussions.
This work was supported by the German DFG
Grant SFB/TRR-55, the Spanish 
Grants FPA2010-16963 and FPA2011-25948, the Catalan Grant SGR2009-00894
and the EU ITN STRONGnet 238353.
C.B.\ was also supported by the Studienstiftung des deutschen
Volkes and by the Daimler und Benz Stiftung.
The computations were performed on Regensburg's
iDataCool cluster
and at the Leibniz Supercomputing Centre in Munich.
\end{acknowledgments}

\end{document}